 \documentclass[prc,amsmath,amsfonts,showpacs,eqsecnum,floatfix,twocolumn]{revtex4}

 \usepackage{graphicx}  
 \usepackage{pstricks}  
 \usepackage{bm}

\begin{document}
\hyphenation{Rij-ken}
\hyphenation{Nij-me-gen}
 
\title{
    Extended-soft-core Baryon-Baryon Model \\   
    I. Nucleon-Nucleon Scattering (ESC04)} 
\author{Th.A.\ Rijken}
\affiliation{Institute for Mathematics, Astrophysics, and Particle Physics, \\   
         Radboud University, Nijmegen, The Netherlands } 
 
\date{version of: \today}

\begin{abstract}                                       
The {\it NN} results are presented from the extended-soft-core ({\rm ESC}) 
interactions.
They consist of local- and non-local-potentials due to (i) one-boson-exchanges 
({\rm OBE}), which are the members of nonets of pseudoscalar-, vector-, scalar-, and
axial-mesons, (ii) diffractive-exchanges, (iii) two-pseudoscalar-exchange (PS-PS),
and (iv) meson-pair-exchange ({\rm MPE}). 
We describe a fit to the {\it pp}- and {\it np}-data for 
$0 \le T_{\rm lab} \leq 350$ MeV, having a typical $\chi^2/N_{\rm data}=1.155$. 
Here, we used ca 20 quasi-free physical parameters, 
being coupling constants and cut-off masses.
A remarkable feature of the couplings is that we were able to require them to 
follow rather closely the pattern predicted by the $^3P_0$ quark-pair-creation 
({\rm QPC}) model. 
As a result the 11 {\rm OBE}-couplings are rather constrained, i.e. quasi-free.
Also, the deuteron binding energy and the several {\it NN} scattering lengths 
are fitted.
 
\end{abstract}

\maketitle
 

\section{Introduction}                                     
\label{sec:1}
 
In a series of three papers we present the results recently obtained with 
the extended-soft-core (ESC) model \cite{Rij93} for nucleon-nucleon ({\it NN}), 
hyperon-nucleon ({\it YN}), and hyperon-hyperon ({\it YY}) with $S=-2$. 
For {\it NN} \cite{Rij93,RS96a,RS96b,SR97,RPN02} it has been demonstrated that the ESC-model
interactions give an excellent description of the {\it NN}-data. 
Also for {\it YN} the first attempts \cite{Rij99,Rij00} showed that the 
ESC-approach is potentially rather promising to give improvements w.r.t. the 
one-boson-exchange ({\rm OBE}) soft-core models \cite{MRS89,RSY99}.
As compared to the earlier versions of the ESC-model, we introduce in these
papers two innovations. First, we introduce a zero in the form factor of 
the scalar mesons. Secondly, we exploit the exchange of the axial-vector mesons.
In this first paper of the series, we display the recent results fitting 
exclusively the {\it NN}-data, giving the {\it NN}-model presented in this paper ESC04({\it NN}).
In the second paper, henceforth referred to as II \cite{RY04b},  
we report on the results for ${\it NN} \oplus {\it YN}$, 
in a simultaneous fit of the {\it NN}- and {\it YN}-data. This is a novelty w.r.t. our 
procedure described in previous publications on the Nijmegen work. The advantages
will be discussed in II.
In the third paper, henceforth referred to as III \cite{RY04c},  
we will report on the predictions for {\it YN} and {\it YY} with $S=-2$.         
 

A general modern theoretical framework for the soft-core
interactions is provided by the so called standard-model (SM).
Starting from the SM we consider the stage where the heavy quarks are 
integrated out, leaving an effective QCD-world for the u,d,s quarks. 
The generally accepted scenario is now that the
QCD-vacuum is unstable for momentum transfers for which 
$Q^2 \leq \Lambda_{\chi SB}^2 \approx 1$ GeV$^2$ \cite{Geo93}, causing 
spontaneous chiral-symmetry breaking ($\chi$SB).
A phase-transition of the vacuum generates constituent quark masses
via $\langle 0|\bar{\psi}\psi|0\rangle \neq 0$, and thereby the gluon coupling
$\alpha_s$ is reduced substantially.
In view of the small pion mass, the Nambu-Goldstone bosons associated with
the spontaneous $\chi SB$ are naturally identified with the pseudoscalar
mesons. Also, as a result of the phase-transition the dominating 
degrees of freedom are the baryons and mesons.
In this context, low-energy baryon-baryon interactions are described naturally 
by meson-exchange using form factors at the meson-baryon vertices. 
This way, the phase transition has transformed the effective QCD-world into
an effective hadronic-world. To reduce this complex world with its numerous
degrees of freedom, we consider a next step. 
This is, envisioning the integrating out of the heavy mesons and baryons using
a renormalization procedure a la Wilson \cite{Pol84}, we restrict ourselves to
mesons with $M \leq 1$ GeV$/c^2$, arriving at a so-called 
{\it effective field theory} as the proper arena to describe low energy baryon-baryon
scattering.  This is the general physical basis for the Nijmegen soft-core models.

Because of the composite nature
of the mesons in QCD, the proper description of meson-exchange 
is quite naturally in terms of Regge-trajectories. 
For example, in the Bethe-Salpeter approach to the $Q \bar{Q}$-system any
reasonable interaction leads to Regge poles.  Therefore, in the Nijmegen 
soft-core approach meson-exchange is treated as the dominant part of the
mesonic reggeon-exchange. This includes also the $J=0$ contributions
from the tensor trajectories ($f_2$,$f_2^\prime$ and $A_{2}$).
In elastic scattering we notice that the most important
exchange at higher energies is pomeron-exchange. Therefore in 
the soft-core {\rm OBE}-models \cite{NRS78} the
traditional {\rm OBE}-model was extended by including the pomeron, and 
the pomeron parameters determined from the low-energy {\it NN}-data were in 
good agreement with those found at high energy. This feature is also found 
to persist in the ESC-models.
For a more elaborate discussion of the pomeron, and its importance for
the implementation of chiral-symmetry in the soft-core models, 
we refer to \cite{MRS89,Pad89}.
 
The dynamics in the ESC-model is constructed employing the following mesons  
together with {\it flavor} {\rm SU(3)}-symmetry:
\begin{enumerate}
\item   The pseudoscalar-meson nonet $\pi,\ \eta,\ \eta',\ K$ with the
        $\eta-\eta'$ mixing angle $\theta_{P}=-23.0^{0}$
        from the \mbox{Gell-Mann-Okubo} mass formula.
\item   The vector-meson nonet $\rho,\ \phi,\ K^{\star},\ \omega$ with
        the $\phi-\omega$ ideal mixing angle $\theta_{V}= 37.56^{0}$.
\item   The axial-vector-meson nonet $a_1, f_1\ K_1, f_1'$ with
        the $f_1-f_1'$ mixing angle $\theta_{A}=  47.3^{0}$ \cite{SR97}.
\item   The scalar-meson nonet 
        $a_0(962)=\delta,\ f_0(993)=S^{\star},\ \kappa,f_0(760)=\varepsilon$
        with a free $S^{\star}-\varepsilon$ mixing angle $\theta_{S}$
        to be determined in a fit to the {\it YN}-data.
\item   The `diffractive' contribution from the pomeron P, and the tensor-mesons
        $f_2$, $f_2^\prime$, and $A_{2}$.
        These interactions will give mainly repulsive contributions 
        of a gaussian type to the potentials in all channels. 
        In the present ESC-model we have taken $g_{A_2}=g_{BBf_2}=g_{BBf_2'}=0$, 
        i.e. only the pomeron contributes.
\end{enumerate}
 
The BBM-vertices are described by: (i) coupling constants and $F/(F+D)$-ratio's
obeying broken flavor {\rm SU(3)}-symmetry, see paper II for details, 
and (ii) gaussian form factors. 
This type of form factor is like the often used residue functions in Regge phenomenology.
Also, from the point of view of the (nonrelativistic) quark models
a gaussian behavior of the form factors is most natural. 
Here, we remark that in the ESC-models the two-meson-cut contributions to the 
form factors are taken into account using meson-pair exchanges ({\rm MPE}) (see below).
Evidently, with cut-off 
masses $\Lambda \approx 1$ GeV, these form factors assure a soft behavior of the potentials 
in configuration space at small distances. The form factors depend on the 
{\rm SU(3)} assignment of the mesons, as described in detail in \cite{RSY99}.

The potentials of the ESC-model are generated by 
\begin{enumerate}
\item [(i)] \underline{One-boson-exchange ({\rm OBE})}.
The treatment of the {\rm OBE} in the soft-core approach has been given for 
 {\it NN} in \cite{NRS78}, and for {\it YN} in \cite{MRS89}. With respect to these
{\rm OBE}-interactions the present ESC-model contains, as mentioned above, two innovations. 
First, in the scalar meson form-factor we have introduced a zero. This zero is natural 
in the $^3P_0$-pair-creation ({\rm QPC}) \cite{Mic69,Yao73,Yao75}  model for the coupling 
of the mesonic quark-antiquark 
($Q\bar{Q}$) system to baryons. The scalar meson, being itself in this picture
a $^3P_0$ $Q\bar{Q}$-bound state, gets a zero when it couples to a baryon.
A pragmatic reason to exploit such a zero is that in this way we were able to 
avoid a bound state in $\Lambda N$-scattering. Secondly, for the first time 
we incorporated axial-meson exchange in the potentials. As is well known, they
are considered as the chiral partners of the vector mesons. It turned out that 
the strength of the axial-meson exchanges is found to agree with the theoretical 
determination $g_{a_1} \approx (m_{a_1}/m_\pi) f_{NN\pi}$ \cite{Sch68}.
\item [(ii)] \underline{ Two-meson-exchange (TME)}.
The configuration space soft-core uncorrelated two-meson exchange for {\it NN} 
has been derived 
in \cite{Rij91,RS96a}. We use these potentials in this paper for PS-PS exchange.
Here, we give a complete {\rm SU(3)}-symmetry treatment in {\it NN}, 
as well as in {\it YN} and {\it YY}. For 
example, we include double $K$-exchange in {\it NN}-scattering.
Similarly in papers II and III their generalization to {\it YN} respectively {\it YY}.
The PS-PS potentials contain the important long-range two-pion potentials. The other kind
of two-meson exchange, as pseudoscalar-vector (PS-V), 
and pseudoscalar-scalar (PS-S) etc.
are supposed to be less important, because of cancellations, 
and can be covered by {\rm OBE} in an effective manner.
Of course, this gives some contamination in the meson-baryon coupling constants.
\item [(iii)] \underline{Meson-pair-exchange ({\rm MPE})}.
These have been described for {\it NN} and justified in \cite{RS96b}. 
Again, in II and III the generalization is used in {\it YN} and {\it YY}. 
Also, the treatment given is complete as far as {\rm SU(3)} is concerned.
In \cite{RS96b,SR97} it is argued that the {\rm MPE}-potentials are thought 
to represent effects of heavy meson-exchange as well as meson-baryon resonances. 
Here we in particularly think about the $\pi N$ resonances, like $\Delta_{33}$.  
\end{enumerate}

A remarkable achievement with the ESC-model, in the version as described above, is that
for the first time we could constrain the 
NNM-couplings such that they are close to the predicted values of the {\rm QPC}-model.
With the same parameters for the quark-model, we find relations like $g_\epsilon 
\approx g_\omega \approx 3 g_\rho \approx 3 g_{a_0}$. Moreover, with the same 
$^3P_0$-parameters the predicted $g_{a_1}$ agrees well with that of \cite{Sch68}.

A particular new feature of these new ESC-models is that we can allow for 
{\rm SU(3)}-symmetry
breaking of the coupling constants. In this breaking it is assumed that the amplitude
for the creation of strange quarks from the vacuum is different than for non-strange 
quarks. We consider this possibility explicitly in paper II, but in this paper
we will assume, apart from meson-mixing, not such an {\rm SU(3)}-breaking.

The contents of this paper is as follows. In section II\ we review
the definition of the ESC-potentials in the context of the
relativistic two-body equations, the Thompson-, and Lippmann-Schwinger-equation.
Here, we exploit the Macke-Klein \cite{Klein53} framework in Field-Theory.
For the Lippmann-Schwinger equation we introduce the usual potential forms
in Pauli spinor space. We include here the central ($C$), the spin-spin
($\sigma$), the tensor ($T$), the spin-orbit ($SO$), the quadratic
spin-orbit ($Q_{12}$), and the antisymmetric spin-orbit ($ASO$)
potentials. For TME-exchange, in the approximations made in \cite{RS96a,RS96b}
only the central, spin-spin, tensor, and spin-orbit potentials occur.
In section III\ the ESC-potentials in momentum space are given, emphasizing the differences
with earlier publications on the soft-core interactions. We discuss the {\rm OBE}-potentials, 
the PS-PS-interactions, and the {\rm MPE}-interactions.
In section IV\ we discuss  the coupling constants from the point of view of the $^3P_0$-model.
In section V\ the {\it NN} results are displayed for coupling constants, scattering phases, 
low-energy parameters, and deuteron properties.
Finally in section VI\ we give a general discussion and outlook.                   

Appendix A contains the derivation of the axial-meson exchange potentials. 
 

\section{Two-Body Integral Equations in Momentum Space}

\subsection{Relativistic Two-Body Equations}
We consider the nucleon-nucleon reactions 
\begin{eqnarray}
 N(p_{a},s_{a})+N(p_{b},s_{b}) \rightarrow
    N(p_{a'},s_{a'})+N(p_{b'},s_{b'}) &&
\label{eq:20.1} \end{eqnarray}
with the total and relative four-momenta for the initial
and the final states
\begin{equation}
\begin{array}{lcl}
 P = p_{a} + p_{b} &,& P' = p_{a'} + p_{b'}\ , \\
 p = \frac{1}{2}(p_{a}-p_{b}) &,& p' = \frac{1}{2}(p_{a'}-p_{b'})\ ,
\end{array} 
\label{eq:20.2} \end{equation}
which become in the center-of-mass system (cm-system) for a and b
on-mass-shell
\begin{equation}
 P = ( W , {\bf 0}) \hspace{0.2cm} , \hspace{0.2cm} p = ( 0 , {\bf p})
 \hspace{0.2cm} , \hspace{0.2cm} p' = ( 0 , {\bf p}') \ .
\label{eq:20.3a} \end{equation}
In general, the particles are off-mass-shell in the Green-functions.
In the following of this section, the on-mass-shell momenta for the initial
and final states are denoted respectively by ${\bf p}$ and ${\bf p}'$.
So, $p_{a}^{0}=E_{a}({\bf p})=\sqrt{{\bf p}^{2}+M_{a}^{2}}$ and
$p_{a'}^{0}=E_{a'}({\bf p}')=\sqrt{{\bf p'}^{2}+M_{a'}^{2}}$, and
similarly for b and b'.
Because of translation-invariance $P=P'$ and
$W=W'=E_{a}({\bf p})+E_{b}({\bf p})=E_{a'}({\bf p}')+E_{b'}({\bf p}')$.
The two-particle states we normalize in the following way
\begin{eqnarray}
  \langle {\bf p}_{1}',{\bf p}_{2}'|{\bf p}_{1},{\bf p}_{2}\rangle
  &=& (2\pi)^{3}2E({\bf p}_{1})
  \delta^{3}({\bf p}_{1}'-{\bf p}_{1})\cdot \nonumber\\
  && \times (2\pi)^{3}2E({\bf p}_{2})
  \delta^{3}({\bf p}_{2}'-{\bf p}_{2})\ .
\label{eq:20.3b} \end{eqnarray}
 
The relativistic two-body scattering-equation for the scattering amplitude 
reads \cite{Feyn49,Schw51,SB51}
\begin{eqnarray}
    M(p',p;P) &=& I(p',p;P) + 
 \int\!d^{4}p^{\prime\prime}\; I(p',p^{\prime\prime};P)\cdot \nonumber\\         
 && \times G(p^{\prime\prime};P)\ M(p^{\prime\prime},p;P)\  ,
\label{BSeq} \end{eqnarray}
where $M(p',p;P)$ is a $16\times 16$-matrix in Dirac-space, and 
the contributions to the kernel $I(p,p')$ come from the two-nucleon-irreducible
Feynman diagrams.
In writing (\ref{BSeq}) we have taken out an overall $\delta^4(P'-P)$-function
and the total four-momentum conservation is implicitly understood henceforth.
 
The two-baryon Green function $G(p;P)$ in (\ref{BSeq}) is simply the
product of the free propagators for, in general, the baryons of line (a) and (b).
The baryon Feynman propagators are given 
by the well known formula
\begin{eqnarray}
    G^{(s)}_{\{\mu\},\{\nu\}}(p) &=& \int d^{4}x\ \langle 0|T(
    \psi^{(s)}_{\{\mu\}}(x)\bar{\psi}^{(s)}_{\{\nu\}}(0))|0\rangle\
    e^{i p\cdot x} \nonumber \\[0.2cm]
  & = & \frac{\Pi^{s}(p)}{p^{2}-M^{2}+i\delta}
\label{Greens} \end{eqnarray}
where $\psi^{(s)}_{\{\mu\}}$ is the free Rarita-Schwinger field
which describes the nucleon $(s=\frac{1}{2})$, the $\Delta_{33}$-resonance 
$(s=\frac{3}{2})$, etc. (see for example \cite{Car71}). 
For the nucleon, the only case considered in this paper,  $\{\mu\}= \emptyset$
and for e.g. the  $\Delta$-resonance $\{\mu\}= \mu$. For the rest of this 
paper we deal only with nucleons.\\
In terms of these one-particle Green-functions
the two-particle Green-function in (\ref{BSeq}) is
\begin{eqnarray}
    G(p ; P) &=&  \frac{i}{(2\pi)^{4}}
 \left[ \frac{ \Pi^{(s_{a})}(\frac{1}{2} P+p)}
 { ( \frac{1}{2} P + p)^{2} - M_{a}^{2}+i\delta} \right]^{(a)}\cdot \nonumber\\ 
 && \times \left[ \frac{ \Pi^{(s_{b})}(\frac{1}{2} P-p)}
 { ( \frac{1}{2} P - p)^{2} - M_{b}^{2}+i\delta} \right]^{(b)}\ .
\label{Green} \end{eqnarray}
 
Using now a complete set of on-mass-shell spin s-states in the first
line of (\ref{Greens}) one finds that the Feynman propagator of a
spin-s baryon off-mass-shell can be written as \cite{BD65}
\begin{eqnarray}
   \frac{ \Pi^{(s)}(p) }{p^{2} -M^{2}+ i \delta} &=&
   \frac{M}{E({\bf p})}
 \left[\frac{\Lambda_{+}^{(s)}({\bf p})}{p_{0}-E({\bf p})+i\delta} 
 \right.\nonumber\\ &&\left. \hspace{1cm} 
 -\frac{\Lambda_{-}^{(s)}(-{\bf p})}{p_{0}+E({\bf p})-i\delta}\right]\ ,
\label{eq:20.7} \end{eqnarray}
 for $s= \frac{1}{2}, \frac{3}{2}, \ldots $.
Here, $\Lambda_{+}^{(s)}({\bf p})$ and $\Lambda_{-}^{(s)}({\bf p})$ are the
on-mass-shell projection operators on the positive- and
negative-energy states. For the nucleon 
they are 
\begin{eqnarray}
    \Lambda_{+}({\bf p}) &=& \sum_{\sigma =-1/2}^{+1/2}
    u({\bf p},\sigma)\otimes\bar{u}({\bf p},\sigma)\ , \nonumber\\
    \Lambda_{-}({\bf p}) &=& -\sum_{\sigma =-1/2}^{+1/2}
    v({\bf p},\sigma)\otimes\bar{v}({\bf p},\sigma)\ ,                      
\label{projop1} \end{eqnarray} 
where $u({\bf p},\sigma)$ and $v({\bf p},\sigma)$ are the Dirac spinors for
spin-$1/2$ particles, and $E({\bf p})=\sqrt{{\bf p}^2+M^2}$
with $M$ the nucleon mass.
Then, in the cm-system, where ${\bf P}=0$ and $P_{0}=W$,
the Green-function can be written as

 \begin{widetext}

\begin{eqnarray}
 G(p;W) &=& \frac{i}{(2\pi)^4}
  \left(\frac{M_a}{E_a({\bf p})} \right)
   \left[ \frac{\Lambda_+^{(s_a)}({\bf p})}
{\frac{1}{2} W+p_0-E_a({\bf p})+i\delta}
    -\frac{\Lambda_-^{(s_a)}(-{\bf p})}
    {\frac{1}{2} W+p_0+E_a({\bf p})-i\delta}\right] \nonumber \\[0.4cm]
& & \hspace*{0.7cm} \times
  \left(\frac{M_b}{E_b({\bf p})}\right)
 \left[ \frac{\Lambda_+^{(s_b)}(-{\bf p})}
 {\frac{1}{2} W-p_0-E_b({\bf p})+i\delta}
    - \frac{\Lambda_-^{(s_b)}( {\bf p})}
    {\frac{1}{2} W-p_0+E_b({\bf p})-i\delta}\right]
 \label{prop1} \end{eqnarray} 


Multiplying out (\ref{prop1})
we  write the ensuing terms in shorthand notation
\begin{eqnarray}
    G(p;W) &=& G_{++}(p;W) + G_{+-}(p;W) 
 + G_{-+}(p;W) + G_{--}(p;W)\ ,
\label{eq:20.10} \end{eqnarray}
where $G_{++}$ etc. corresponds to the term with 
$\Lambda_{+}^{s_{a}} \Lambda_{+}^{s_{b}}$ etc.
Introducing the spinorial amplitudes                    
\begin{equation}
 M_{r's';rs}(p',p;P)= \bar{u}^{r'}(p'_a,s'_a) \bar{u}^{s'}(p'_b,s'_b)
 M(p',p;P)\ u^{r}(p_a,s_a)\ u^{s}(p_b,s_b)\ ,
 \hspace{0.4cm}  (r,s=+,-) \ ,
\label{eq:20.11} \end{equation}
with $(r,s)=+$ for the positive energy Dirac spinors, and $(r,s)=-$ for the  
negative energy ones.     
Then, the two-body equation, (\ref{BSeq}) for the spinorial amplitudes becomes
 \begin{eqnarray}
     M_{r's';rs}(p',p;P) &=& I_{r's';rs}(p',p;P) + 
  \sum_{r'',s''} \int\! d^{4}p^{\prime\prime}\ 
  I_{r's';r^{\prime\prime} s^{\prime\prime}}(p',p^{\prime\prime};P)\cdot 
 \nonumber \\ && \times 
  G_{ r^{\prime\prime} s^{\prime\prime} } (p^{\prime\prime} ; P)\ 
  M_{r^{\prime\prime} s^{\prime\prime} ;rs}(p^{\prime\prime},p;P)\ .
 \label{eq:20.12} \end{eqnarray}
 
Invoking `dynamical pair-suppression', as discussed in
\cite{Rij91}, (\ref{eq:20.12}) reduces to a $4\times 4$-dimensional equation
for $M_{++;++}$, {\it i.e.}
\begin{eqnarray}
    M_{++;++}(p',p;P) &=& I_{++;++}(p',p;P) + \int\!d^{4}p^{\prime\prime}\; 
 I_{++;++}(p',p^{\prime\prime};P)\cdot \nonumber\\         
 && \times G_{++}(p^{\prime\prime};P)\ M_{++;++}(p^{\prime\prime},p;P)\  ,
\label{eq:20.13} \end{eqnarray}
with the Green-function 
\begin{eqnarray}
 G_{++}(p;W)&=& \frac{i}{(2\pi)^{4}}
 \left[\frac{M_{a}M_{b}}{E_{a}({\bf p})E_{b}({\bf p})}\right]
 \cdot     \left[\frac{1}{2} W+p_{0}-E_{a}({\bf p})+i\delta\right]^{-1} 
\left[\frac{1}{2} W-p_{0}-E_{b}({\bf p})+i\delta\right]^{-1}\ .
\nonumber\\
 \label{Greenth} \end{eqnarray}


\subsection{Three-Dimensional Equation}
In \cite{Rij91} we introduced starting from the Bethe-Salpeter 
equation for the two-baryon wave function $\psi(p^\mu)$ and applying the 
Macke-Klein procedure \cite{Klein53}. In this paper we employ the same 
procedure, but now for the two-baryon scattering amplitude $M(p',p;P)$.
For any function $f(p_1,\ldots,p_n)$ we define the projection \cite{Klein74}
\begin{eqnarray}
 P_{R,p_i} f(p_1,\ldots,p_n) &=& f(p_1,\ldots,p_n) P_{L,i}
 \equiv \oint_{UHP} dp_{i,0}\ A_W(p_i)\ f(\ldots, p_i,\ldots)\ , 
 \label{eq:20.15} \end{eqnarray}
where the contour consists of the real axis and the infinite semicircle in the 
upper half plane (UHP), and with Macke's right-inverse of the $\int dp_0$ operation
\begin{eqnarray}
 A_W(p) &=& (2\pi i)^{-1} \left(\frac{1}{p_0+E_p-W-i\delta} + 
                                \frac{1}{-p_0+E_p-W-i\delta}\right) \nonumber\\
  &=& -\frac{1}{2\pi i} \frac{W-{\cal W}({\bf p})}            
  {F^{(a)}_{W}({\bf p},p_{0}) F^{(b)}_{W}(-{\bf p},-p_{0})}\ . 
 \label{eq:20.16} \end{eqnarray}
Here, we used the frequently used notations           
\begin{eqnarray}
  &&  F_{W}({\bf p},p_{0}) = p_{0}-E({\bf p})+ \frac{1}{2} W + i\delta\ \ ,\ \
      {\cal W}({\bf p}) = E_{a}({\bf p}) + E_{b}({\bf p})\ .
\label{eq:20.17} \end{eqnarray}
Notice that the Green function (\ref{Greenth} can be written as
\begin{eqnarray}
 G_{++}(p;W)&=& \frac{1}{(2\pi)^{3}}
 \left[\frac{M_{a}M_{b}}{E_{a}({\bf p})E_{b}({\bf p})}\right]\
 A_W(p)\left( W - {\cal W}(p)+i\delta\right)^{-1}\ .
\label{eq:20.18} \end{eqnarray}

Now, we make the rather solid assumption that for the scattering amplitudes,
the UHP contains no poles or branch points in the $p_0$-variable. 
Then, one sees from (\ref{eq:20.15}) that as a result 
of the  $P_{R,p_i}$-operation the argument $p_{i0} \rightarrow W-E({\bf p}_i)$, and 
similarly for $P_{L,p_i}$. Introducing the projections
\begin{subequations}
\begin{eqnarray}
 P_{R,p'} M_{++;++}(p',p;P)\ P_{L,p} &\equiv& M({\bf p}',{\bf p}|W)\ , 
 \\ 
 P_{R,p'} I_{++;++}(p',p;P)\ P_{L,p} &\equiv & K^{irr}({\bf p}',{\bf p}|W)\ ,      
\label{eq:20.19} \end{eqnarray}
\end{subequations}
we apply this to equation (\ref{eq:20.13}). This gives
\begin{eqnarray}
 && \hspace{-5mm}
 M({\bf p}',{\bf p}| W) = K^{irr}({\bf p}',{\bf p}|W) + 
 \int\!\frac{d^{3}p^{\prime\prime}}{(2\pi)^3}\ 
 \left[\frac{M_{a}M_{b}}{E_{a}({\bf p}^{\prime\prime})E_{b}({\bf p}^{\prime\prime})}\right]\
 \left( W-{\cal W}({\bf p}^{\prime\prime})+i\delta\right)^{-1}\cdot  
 \nonumber\\ && \hspace{-5mm} \times
 \left\{\int^\infty_{-\infty}dp_0^{\prime\prime} 
 \left.I_{++;++}(p',p^{\prime\prime};P)\right|_{p_0'=W-E({\bf p}')}\         
 A_W(p^{\prime\prime})\ 
 \left.M_{++;++}(p^{\prime\prime},p;P)\right|_{p_0=W-E({\bf p})}\right\} ,
\label{eq:20.20} \end{eqnarray}
Next, we redefine $ M({\bf p}^{\prime\prime},{\bf p}|W)$ by 
\begin{eqnarray}
M({\bf p}^{\prime},{\bf p}|W) &\rightarrow&
 \sqrt{\frac{M_a M_b }{E_a({\bf p}') E_b({\bf p}')} }\
 M({\bf p}^{\prime},{\bf p}|W)\              
 \sqrt{\frac{M_a M_b }{E_a({\bf p}) E_b({\bf p})} }\ ,
\label{eq:20.21} \end{eqnarray}
and similarly for $K^{irr}({\bf p}^{\prime\prime},{\bf p}|W)$.   
The thus redefined quantities obey again equation (\ref{eq:20.20}), except for the 
factor $\left[\ldots\right]$ on the right-hand side.
Closing now the contour of the $p_0^{\prime\prime}$-integration in the upper-half plane, 
one picks up again only the contribution at 
$p_0^{\prime\prime}= W-E({\bf p}^{\prime\prime})$,
which means that (\ref{eq:20.20}) becomes the Thompson equation \cite{Thom70}
\begin{eqnarray}
 && \hspace{-1cm} M({\bf p}',{\bf p}| W) = K^{irr}({\bf p}',{\bf p}|W) + 
 \int\!\frac{d^{3}p^{\prime\prime}}{(2\pi)^3} 
 K^{irr}({\bf p}',{\bf p}^{\prime\prime}|W)\ E_{2}^{(+)}({\bf p}^{\prime\prime}; W)\
 M({\bf p}^{\prime\prime},{\bf p}|W)\ ,  
\label{eq:20.22} \end{eqnarray}
where 
$ E_{2}^{(+)}({\bf p}^{\prime\prime}; W)=  
 \left( W-{\cal W}({\bf p}^{\prime\prime})+i\delta\right)^{-1} $.  
Written explicitly, we have from (\ref{eq:20.19}) that the two-nucleon irreducible
kernel is given by
\begin{eqnarray}
  K^{{\it irr}}({\bf p}',{\bf p}| W)&=& -\frac{1}{(2\pi)^{2}}
 \sqrt{\frac{M_{a}M_{b}}{E_{a}({\bf p}') E_{b}({\bf p}')} }
 \sqrt{\frac{M_{a}M_{b}}{E_{a}({\bf p}) E_{b}({\bf p})} }
 \left(W-{\cal W}({\bf p}')\right)\left(W-{\cal W}({\bf p})\right)
 \nonumber \\[0.2cm] &\times&
  \int_{-\infty}^{+\infty} dp_{0}'
   \int_{-\infty}^{+\infty} dp_{0} \left[ \vphantom{\frac{A}{A}}
 \left\{F_{W}^{(a)}({\bf p}',p_{0}')
 F_{W}^{(b)}(-{\bf p}',-p_{0}')\right\}^{-1}
 \right.\nonumber \\[0.2cm] &\times& \left.
 \left[ I(p_{0}',{\bf p}'; p_{0},{\bf p}) \right]_{++,++}
    \left\{F_{W}^{(a)}({\bf p},p_{0})
    F_{W}^{(b)}(-{\bf p},-p_{0})\right\}^{-1}
\vphantom{\frac{A}{A}}\right]\ , 
\label{Thomp2}  \end{eqnarray}
which is the same expression as we exploited in our previous papers, e.g.
\cite{Rij91,RS96a,RPN02}. In the latter we exploited the three-dimensional  
wave function according to Salpeter \cite{Salp52} combined with the Macke-Klein
ansatz \cite{Klein53}. For the scattering amplitude the derivation given above 
is more direct. For a discussion and comparison with other three-dimensional 
reductions of the Bethe-Salpeter equation we refer to \cite{Klein74}.
In case one does not assume the strong pair-suppression, one must study instead 
of equation (\ref{eq:20.13}) a more general equation with couplings between the 
positive and negative energy spinorial amplitudes. Also to this more general case
one can apply the described three-dimensional reduction, and we refer the reader to 
\cite{Klein74} for a treatment of this case.



 The $M/E$-factors in (\ref{Thomp2}) are due to the difference
between the relativistic and the non-relativistic normalization of
the two-particle states. In the following we simply put
$M/E({\bf p})=1$ in the kernel $K^{irr}$ Eq.~(\ref{Thomp2}). The corrections
to this approximation would give $(1/M)^{2}$-corrections
to the potentials, which we neglect in this paper. In the same approximation
there is no difference between the Thompson \cite{Thom70}
and the Lippmann-Schwinger equation, when the connection between these 
equations is made using multiplication factors. Henceforth, we will not
distinguish between the two.
 
The contributions to the two-particle irreducible kernel
$K^{{\it irr}}$ up to second order in the meson-exchange
are given in detail in \cite{RS96a,RS96b}.

 \end{widetext}

 
\subsection{Lippmann-Schwinger Equation}

 \begin{figure}   
 {\includegraphics[3.5in,7in][5.5in,9.6in]{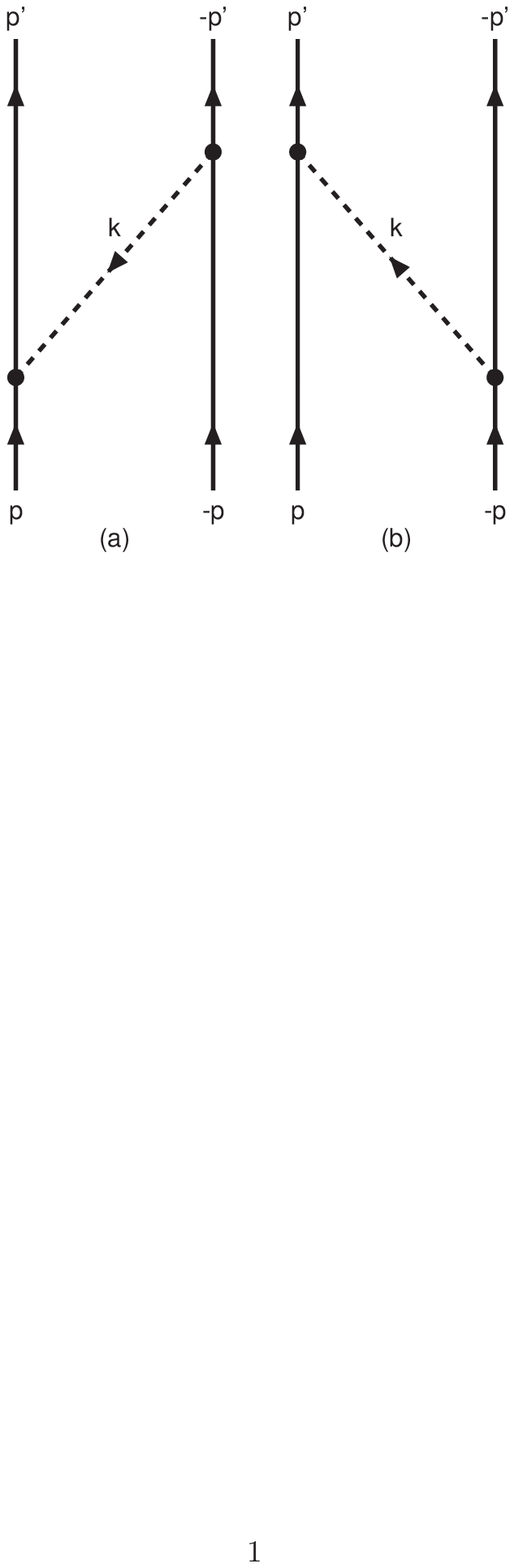}}
\caption{One-boson-exchange graphs: 
        The dashed lines with momentum ${\bf k}$ refers to the
        bosons: pseudoscalar, vector, axial-vector, or scalar mesons.}
\label{obefig}
 \end{figure}

 \begin{figure}   
 {\includegraphics[3.2in,5in][5.2in,10in]{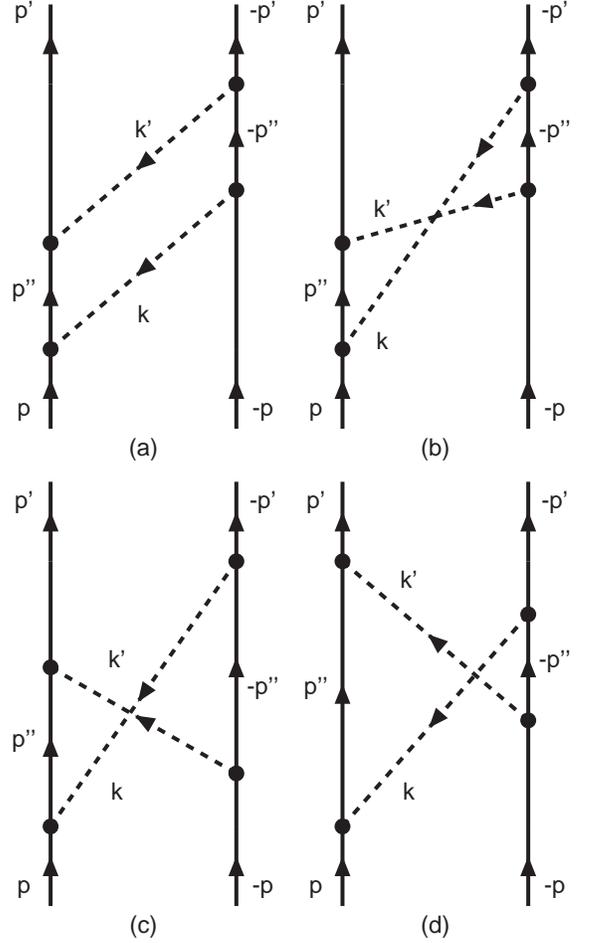}}
\caption{BW two-meson-exchange graphs: (a) planar and (b)-(d) crossed
        box. The dashed line with momentum ${\bf k}$ refers to the
        pion and the dashed line with momentum ${\bf k}^\prime$ refers
        to one of the other (vector, scalar, or pseudoscalar) mesons.
        To these we have to add the ``mirror'' graphs, and the
        graphs where we interchange the two meson lines.}
\label{bwfig}
 \end{figure}

 \begin{figure}   
 {\includegraphics[3.2in,7.5in][5.2in,9.8in]{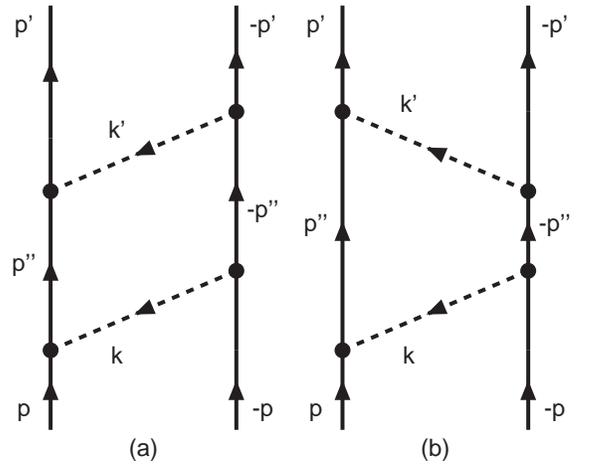}}
\caption{Planar-box TMO two-meson-exchange graphs.
         Same notation as in Fig.~\protect\ref{bwfig}.
         To these we have to add the ``mirror'' graphs, and the
         graphs where we interchange the two meson lines.}
 \label{tmofig}
 \end{figure}

The transformation of (\ref{eq:20.22}) to the Lippmann-Schwinger 
equation can be effectuated by defining
\begin{subequations}
\begin{eqnarray}
 T({\bf p}',{\bf p}) &=& N({\bf p}')\ M({\bf p}',{\bf p}|W)\ N({\bf p})\ , 
 \\[0.2cm]
 V({\bf p}',{\bf p}) &=& N({\bf p}')\ K^{irr}({\bf p}',{\bf p}|W)\ N({\bf p})\ ,       
\label{eq:20.24} \end{eqnarray}
\end{subequations}
where the transformation function is 
\begin{equation} 
 N({\bf p}) = \sqrt{\frac{{\bf p}_i^2-{\bf p}^2}{2M_N(E\left({\bf p}_i)-E({\bf p})\right)}}\ .
\label{eq:20.25} \end{equation}
Application of this transformation, yields the Lippmann-Schwinger
equation
\begin{eqnarray}
    T({\bf p}',{\bf p}) &=& V({\bf p}',{\bf p}) +
    \int \frac{d^{3}p''}{(2\pi)^3} \nonumber\\ 
 && \times V({\bf p}',{\bf p}'')\ g({\bf p}''; W)\; T({\bf p}'',{\bf p})
\label{eq:20.26} \end{eqnarray}
with the standard Green function     
\begin{equation}
    g({\bf p};W) = 
    \frac{M_N}{{\bf p}_i^{2}-{\bf p}^{2}+i\delta} \ .
\label{eq:20.27} \end{equation}
The corrections to the approximation $E_{2}^{(+)} \approx g({\bf p}; W)$ 
are of order $1/M^{2}$, which we neglect hencforth.
 
The transition from Dirac-spinors to
Pauli-spinors, is given in Appendix C of \cite{Rij91}, where we write for the 
the Lippmann-Schwinger equation in the 4-dimensional Pauli-spinor space
\begin{eqnarray}
 {\cal T}({\bf p}',{\bf p})&=&{\cal V}({\bf p}',{\bf p}) + \int \frac{d^{3} p''}{(2\pi)^3}\
 \nonumber\\ && \times 
 {\cal V}({\bf p}',{\bf p}'')\  g({\bf p}''; W)\
  {\cal T}({\bf p}'',{\bf p})\ .
 \label{eq:20.28} \end{eqnarray}

The ${\cal T}$-operator in Pauli spinor-space is defined by
\begin{eqnarray}
 && \chi^{(a)\dagger}_{\sigma'_{a}}\chi^{(b)\dagger}_{\sigma'_{b}}\; 
 {\cal T}({\bf p}',{\bf p})\;
 \chi^{(a)}_{\sigma_{a}}\chi^{(b)}_{\sigma_{b}}  =              
 \bar{u}_{a}({\bf p}',\sigma'_{a})\bar{u}_{b}(-{\bf p}',\sigma'_{b}) \nonumber\\
 && \times \tilde{T}({\bf p}',{\bf p})\;
     u_{a}({\bf p},\sigma_{a}) u_{b}(-{\bf p},\sigma_{b}) \ .
 \label{eq:20.29} \end{eqnarray}
and similarly for the ${\cal V}$-operator.
Like in the derivation of the {\rm OBE}-potentials \cite{NRS78,NRS77}
we make off-shell and on-shell the approximation,
  $ E({\bf p})= M + {\bf p}^{2}/2M $
 and $ W = 2\sqrt{{\bf p}_i^{2}+M^{2}} = 2M + {\bf p}_i^{2}/M$ ,     
everywhere in the interaction kernels, which, of course,
is fully justified for low energies only. 
In contrast to these kind of approximations, of course the full
${\bf k}^{2}$-dependence of the form factors is kept
throughout the derivation of the TME. 
Notice that the Gaussian form factors suppress the high momentum
transfers strongly. This means that the contribution to the potentials
from intermediate states which are far off-energy-shell can not
be very large. 

Because of rotational invariance and parity conservation, the ${\cal T}$-matrix, which is
a $4\times 4$-matrix in Pauli-spinor space, can be expanded 
into the following set of in general 8 spinor invariants, see for example 
\cite{SNRV71}. Introducing \cite{notation1}
\begin{equation}
  {\bf q}=\frac{1}{2}({\bf p}'+{\bf p})\ , \
  {\bf k}={\bf p}'-{\bf p}\ , \           
  {\bf n}={\bf p}\times {\bf p}'\ ,
\label{eq:20.30} \end{equation}
with, of course, ${\bf n}={\bf q}\times {\bf k}$,
we choose for the operators $P_{j}$ in spin-space
\begin{subequations}
\begin{eqnarray}
  P_{1} &=& 1\ ,  \\[0.2cm]
 P_{2} &=& \mbox{\boldmath $\sigma$}_1\cdot\mbox{\boldmath $\sigma$}_2\ ,
\\[0.2cm]
 P_{3}&=& (\mbox{\boldmath $\sigma$}_1\cdot{\bf k})(\mbox{\boldmath $\sigma$}_2\cdot{\bf k})
 -\frac{1}{3}(\mbox{\boldmath $\sigma$}_1\cdot\mbox{\boldmath $\sigma$}_2)\
  {\bf k}^2\ , \\[0.2cm]
 P_{4}&=&\frac{i}{2}(\mbox{\boldmath $\sigma$}_1+
 \mbox{\boldmath $\sigma$}_2)\cdot{\bf n}\ , \\[0.2cm]
 P_{5}&=&(\mbox{\boldmath $\sigma$}_1\cdot{\bf n})(\mbox{\boldmath $\sigma$}_2\cdot{\bf n})\ ,
\\[0.2cm]
  P_{6}&=&\frac{i}{2}(\mbox{\boldmath $\sigma$}_1-
 \mbox{\boldmath $\sigma$}_2)\cdot{\bf n}\ ,          
\\[0.2cm]
 P_{7}&=&(\mbox{\boldmath $\sigma$}_1\cdot{\bf q})(\mbox{\boldmath $\sigma$}_2\cdot{\bf k})
 +(\mbox{\boldmath $\sigma$}_1\cdot{\bf k})(\mbox{\boldmath $\sigma$}_2\cdot{\bf q})\ , 
\\[0.2cm]
 P_{8}&=&(\mbox{\boldmath $\sigma$}_1\cdot{\bf q})(\mbox{\boldmath $\sigma$}_2\cdot{\bf k})
 -(\mbox{\boldmath $\sigma$}_1\cdot{\bf k})(\mbox{\boldmath $\sigma$}_2\cdot{\bf q})\ .
\label{eq:20.31} \end{eqnarray}
\end{subequations}
Here we follow \cite{MRS89}, where in contrast to \cite{NRS78},
we have chosen $P_{3}$ to be a purely `tensor-force' operator.
The expansion in spinor-invariants reads
\begin{equation}
 {\cal T}({\bf p}',{\bf p}) = \sum_{j=1}^8\ \widetilde{T}_j({\bf p}^{\prime 2},{\bf p}^2,
 {\bf p}'\cdot{\bf p})\ P_j({\bf p}',{\bf p})\ .
\label{eq:20.32} \end{equation}
Similarly to (\ref{eq:20.32}) we expand the potentials $V$. Again
following \cite{MRS89}, we neglect the potential forms $P_{7}$ and
$P_{8}$, and also the dependence of the potentials on
${\bf k}\cdot{\bf q}$ . Then, the expansion (\ref{eq:20.32}) reads for
the potentials as follows
\begin{equation}
  {\cal V} =\sum_{j=1}^{4}\tilde{V}_{j}({\bf k}\,^{2},{\bf q}\,^{2})\
 P_{j}({\bf k},{\bf q})\ .
\label{eq:20.33} \end{equation}
 
  \begin{figure}   
 {\includegraphics[3.2in,5in][5.2in,10in]{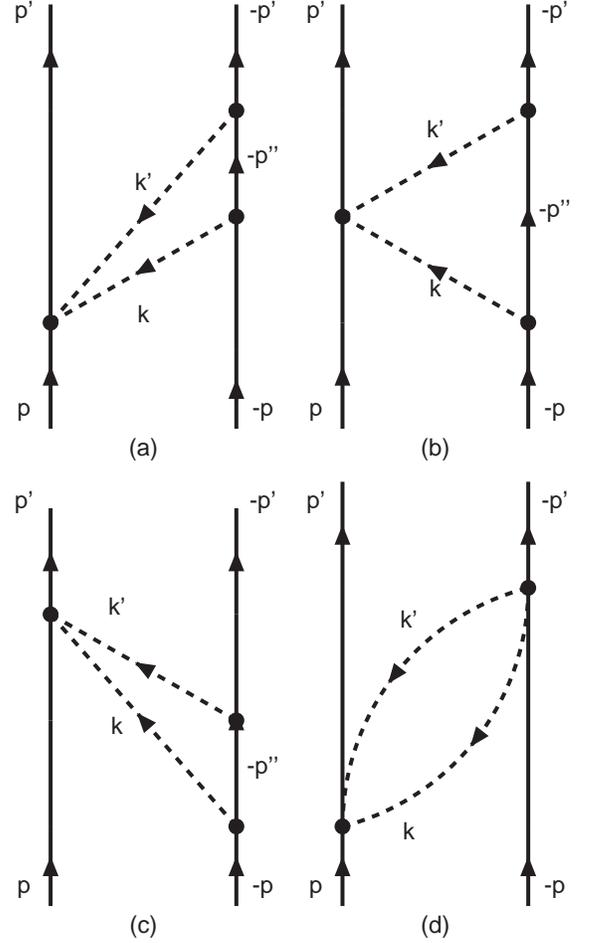}}
 \caption{One- and Two-Pair exchange graphs.
          To these we have to add the ``mirror'' graphs, and the
          graphs where we interchange the two meson lines.}
  \label{pairfig}
  \end{figure}

\section{Extended-Soft-Core Potentials in Momentum Space}
\label{sect.III}
The potential of the ESC-model contains the contributions from 
(i) one-boson-exchanges, Fig.~\ref{obefig}, (ii) uncorrelated 
two-pseudoscalar-exchange, Fig.~\ref{bwfig} and Fig.~\ref{tmofig}, 
and (iii) meson-pair-exchange, Fig~\ref{pairfig}. In this section we 
review the potentials and indicate the changes with respect to 
earlier papers on the {\rm OBE}- and ESC-models.

\subsection{One-Boson-Exchange Interactions in Momentum Space}
\label{sect.IIIa}
The {\rm OBE}-potentials are the 
same as given in \cite{NRS78,MRS89}, with the exception of 
 (i) the zero in the scalar form factor, and 
 (ii) the axial-vector-meson potentials.
Here, we review the {\rm OBE}-potentials briefly, and give those potentials
that are not incuded in the above references.
The local interaction Hamilton densities for the different couplings
are \\ \\
        a) Pseudoscalar-meson exchange
         \begin{equation}
         {\cal H}_{PV}=i\frac{f_{P}}{m_{\pi^{+}}}
         \bar{\psi}\gamma_{\mu}\gamma_{5}
                \psi\partial^{\mu}\phi_{P}\ , \label{eq:3.1}\end{equation}
        b) Vector-meson exchange
       \begin{equation}
   {\cal H}_{V}=ig_{V}\bar{\psi}\gamma_{\mu}\psi\phi_{V}^{\mu}
                +\frac{f_{V}}{4{\cal M}}\bar{\psi}\sigma_{\mu\nu}
                \psi (\partial^{\mu}\phi^{\nu}_{V}-\partial^{\nu}
                      \phi^{\mu}_{V})\ , \label{eq:3.2}\end{equation}
        c) Axial-vector-meson exchange
       \begin{equation}
   {\cal H}_{A}= g_{A}\bar{\psi}\gamma_\mu\gamma_5\psi\phi_{A}^{\mu}
                +\frac{i f_{A}}{{\cal M}}\left[\bar{\psi}\gamma_5       
                \psi\right]\ \partial_{\mu}\phi^{\mu}_{A}\ , \label{eq:3.3}\end{equation}
        We take $f_A=0$, and notice that for the $A_1$-meson the interaction 
        (\ref{eq:3.3}) is part of interaction
        \begin{eqnarray}
        {\cal L}_I^{(A)} &=& 2g_A\left[
   \bar{\psi}\gamma_5\gamma_\mu\frac{\mbox{\boldmath $\tau$}}{2}\psi + 
   \left(\mbox{\boldmath $\pi$}\partial_\mu\sigma
   -\sigma\partial_\mu\mbox{\boldmath $\pi$}\right)
   \right.\nonumber\\ && \left.\vphantom{\frac{A}{A}}
    +f_\pi\partial_\mu \mbox{\boldmath $\pi$}\right]\cdot{\bf A}^\mu\ ,    
        \end{eqnarray}
        which is such that the $A_1$ couples to an almost conserved axial current (PCAC).
        Therefore, the $A_1$-coupling used is compatible with broken 
        ${\rm SU(2)}_V\times {\rm SU(2)}_A$-symmetry \cite{Schw69}.
        
        d) Scalar-meson exchange
         \begin{equation}
               {\cal H}_{S}=g_{S}\bar{\psi}\psi\phi_{S}\ . \label{eq:3.4}\end{equation}
       Here, we used the conventions of \cite{BD65} where 
       $\sigma_{\mu\nu}= [\gamma_{\mu},\gamma_{\nu}]/2i$.
       The scaling masses $m_{\pi^{+}}$ and ${\cal M}$ are 
       chosen to be the charged pion and the proton mass, respectively.
       Note that the vertices for `diffractive'-exchange have the
       same Lorentz structure as those for scalar-meson-exchange.

Including form factors $f({\bf x}'-{\bf x})$ ,
the interaction hamiltonian densities are modified to
\begin{equation}
        H_{X}({\bf x})=\int\!d^{3}x'\,
  f({\bf x}'-{\bf x}){\cal H}_{X}({\bf x}')\ ,
\end{equation}
for $X=PV,\ V$, $A$, $S$, or $D$. Because of the convolutive non-local form, the
potentials in momentum space are the same as for point interactions,
except that the coupling constants are multiplied by the Fourier
transform of the form factors.
 
In the derivation of the $V_{i}$ we employ the same approximations as in 
\cite{NRS78,MRS89}, i.e.
\begin{enumerate}
\item   We expand in $1/M$: 
    $E(p) = \left[ {\bf k}^{2}/4 +
    {\bf q}^{2}+M^{2}\right]^{\frac{1}{2}}$\\
    $\approx M+{\bf k}^{2}/8M + {\bf q}^{2}/2M$
 and keep only terms up to first order in ${\bf k}^{2}/M$ and
 ${\bf q}^{2}/M$. This except for the form factors where
 the full ${\bf k}^{2}$-dependence is kept throughout
 the calculations. Notice that the gaussian form factors
suppress the high ${\bf k}^{2}$-contributions strongly.
\item   In the meson propagators
$       (-(p_{1}-p_{3})^{2}+m^{2})  
        \approx({\bf k}^{2}+m^{2})$ .
\item   When two different baryons are involved at a BBM-vertex
        their average mass is used in the
        potentials and the non-zero component of the momentum transfer
        is accounted for by using an effective mass in
        the meson propagator (for details see \cite{MRS89}).     
\end{enumerate}
 
Due to the approximations we get only a linear dependence on
${\bf q}^{2}$ for $V_{1}$. In the following, we write
\begin{equation}
  V_{i}({\bf k}^{2},{\bf q}^{2})=
  V_{i a}({\bf k}^{2})+V_{i b}({\bf k}^{2}){\bf q}^{2}\ ,
\label{vcdec} \end{equation}
where in principle $i=1,8$. 
 
The {\rm OBE}-potentials are now obtained in the standard way (see e.g.\
\cite{NRS78,MRS89}) by evaluating the {\it BB}-interaction in Born-approximation.
We write the potentials $V_{i}$ of Eqs.~(\ref{eq:20.33}) and
(\ref{vcdec}) in the form
\begin{equation}
  V_{i}({\bf k}\,^{2},{\bf q}\,^{2})=
   \sum_{X} \Omega^{(X)}_{i}({\bf k}\,^{2})
   \cdot \Delta^{(X)} ({\bf k}^{2},m^{2},\Lambda^{2})\ ,
\label{nrexpv2} \end{equation}
 where $X= P,\ V,\ A,\ S$, and $D$ ($P =$ pseudoscalar, $V =$ vector,
 $A=$ axial-vector, $S =$ scalar, and $D =$ diffractive). Furthermore
for $X=P,V$ 
\begin{equation}
   \Delta^{(X)}({\bf k}^{2},m^{2},\Lambda^{2})= e^{-{\bf k}^{2}/\Lambda^{2}}/  
                    \left({\bf k}^{2}+m^{2}\right)\ ,
\label{propm1} \end{equation}
and for $X=S,A$ a zero in the form factor
\begin{equation}
   \Delta^{(S)}({\bf k}^{2},m^{2},\Lambda^{2})= \left(1-{\bf k}^2/U^2\right)\
  e^{-{\bf k}^{2}/\Lambda^{2}}/  
  \left({\bf k}^{2}+m^{2}\right)\ ,
\label{propm2} \end{equation}
and for $X=D$
\begin{equation}
   \Delta^{(D)}({\bf k}^{2},m^{2},\Lambda^{2})=\frac{1}{{\cal M}^{2}}
   e^{-{\bf k}^{2}/(4m_{P}^{2})}\ .
\label{Eq:difdel}
\end{equation}
In the latter expression ${\cal M}$ is a universal
scaling mass, which is again taken to be the proton mass.
The mass parameter $m_{P}$ controls the ${\bf k}^{2}$-dependence of
the pomeron-, $f$-, $f'$-, $A_{2}$-, and $K^{\star\star}$-potentials.

Next, we make remarks which point out the differences in the potentials of this
work as compared to with earlier soft-core model papers:\\

\noindent a)\ For pseudoscalar mesons, the graph's of Fig.~\ref{obefig} give for the 
 second-order 
 potential $ V_{PS}({\bf k},{\bf q}) \approx K^{(2)}_{PS}({\bf p}',{\bf p}|W) $ 
\begin{eqnarray}
 V_{PS}({\bf k},{\bf q}) & = & -\frac{f_{13}f_{24}}{m_\pi^2}\
 \left(1-\frac{({\bf q}^2+{\bf k}^2/4)}{2M_YM_N}\right)\cdot \nonumber\\
 && \hspace{-1.2cm}\times \frac{ (\mbox{\boldmath $\sigma$}_1\cdot{\bf k})
 (\mbox{\boldmath $\sigma$}_2\cdot{\bf k})}{\omega({\bf k})[\omega({\bf k})+a]}\
 \exp\left(-{\bf k}^2/\Lambda^2\right)\ ,
 \label{eq:psx1}\end{eqnarray}
where $a \approx \left({\bf q}^2+{\bf k}^2/4\right)-p_i^2$. Here, $p_i$ is the 
on-energy-shell momentum. On-energy-shell $a=0$, and henceforth we neglect the 
non-adiabatic effects, i.e. $a \neq 0$, in the {\rm OBE}-potentials.
However, we do include the non-local term in (\ref{eq:psx1}, to which we refer 
in the following as the Graz-correction \cite{Graz78}.
From (\ref{eq:psx1}) we find for $\Omega^{(P)}_i$:
\begin{subequations}
     \begin{eqnarray}                       
      \Omega^{(P)}_{2a} & = & g^P_{13}g^P_{24}\left(
      \frac{{\bf k}^{2}}{12M_{Y}M_{N}} \right) \\[0.5cm]
      \Omega^{(P)}_{2b} & = & -g^P_{13}g^P_{24}\left(
      \frac{{\bf k}^{2}}{24M_{Y}^2M_{N}^2} \right) \\[0.5cm]
      \Omega^{(P)}_{3a} & = & -g^P_{13}g^P_{24}\left( \frac{1}{4M_{Y}M_{N}}\right) \\[0.5cm]
      \Omega^{(P)}_{3a} & = & +g^P_{13}g^P_{24}\left( \frac{1}{8M_{Y}^2M_{N}^2}\right)
       \label{eq:psx2}   \end{eqnarray}
\end{subequations}
The $\Omega^{(P)}_{2b,3b}$ contributions were not included in 
\cite{NRS78,MRS89}.\\

\noindent b)\ For vector-, and diffractive {\rm OBE}-exchange we  
refer the reader to Ref.~\cite{MRS89}, where the contributions to the different
$\Omega^{(X)}_{i}$'s for baryon-baryon scattering are given in detail.
Also, it is trivial to obtain from \cite{MRS89} the scalar-meson $\Omega_i$ making the 
substitutions: 
\[
\Omega^{(S)}_{i} \rightarrow \left(1-{\bf k}^2/U^2\right)\ \Omega^{(S)}_{i}\ ,
\]
which now evidently have a zero for ${\bf k}^2 = U^2$.\\
\noindent c)\ For the axial-vector mesons, 
 the detailed derivation of the $\Omega_i^{(A)}$ is given in Appendix~\ref{app:A}.
 Using the approximations (1-5), from the $1^{st}$-term in the axial-meson propagator  
 we get, see  (\ref{eq:A.12}), the following contributions                    
\begin{subequations}
     \begin{eqnarray} 
     \Omega^{(A)}_{2a} & = & -g^A_{13}g^A_{24}  
          \left( 1+\frac{{\bf k}^{2}}{24M_{Y}M_{N}}\right)\ , \\
     \Omega^{(A)}_{2b}&=&
    -g^A_{13}g^A_{24}\  \frac{1}{6M_{Y}M_{N}}\ ,   \\         
     \Omega^{(A)}_{3}&=&
    +g^A_{13}g^A_{24}\  \frac{3}{4M_{Y}M_{N}}\ ,     \\         
     \Omega^{(A)}_{4}&=&
    -g^A_{13}g^A_{24}\  \frac{1}{2M_{Y}M_{N}}\ ,   \\       
     \Omega^{(A)}_{6}&=&
    -g^A_{13}g^A_{24}\ 
     \frac{(M_{N}^{2}-M_{Y}^{2})}{4M_{Y}^2M_{N}^2}\ . 
     \label{eq:axi1} \end{eqnarray}
\end{subequations}
     From the $2^{nd}$-term propagator we get, see (\ref{eq:A.14}),
\begin{subequations}
     \begin{eqnarray}                              
     \hspace{-5mm} \Omega^{(A)}_{2a} & = & -g^A_{13}g^A_{24}  
          \left( 1-\frac{{\bf k}^{2}}{8M_{Y}M_{N}}\right)\cdot
          \frac{{\bf k}^2}{3 m^2}\ ,   \\         
    \hspace{-5mm}  \Omega^{(A)}_{2b}&=& 
    +g^A_{13}g^A_{24}\  \frac{1}{2M_{Y}M_{N}}\cdot
          \frac{{\bf k}^2}{3 m^2}\ ,      \\   
    \hspace{-5mm}  \Omega^{(A)}_{3a}&=&
    -g^A_{13}g^A_{24}  \left(1-\frac{{\bf k}^2}{8M_{Y}M_{N}}\right)\cdot
          \frac{1}{m^2}\ ,   \\
    \hspace{-5mm}  \Omega^{(A)}_{3b}&=&
    +g^A_{13}g^A_{24}\  \frac{1}{2M_{Y}M_{N}}\cdot 
          \frac{1}{m^2}\ .
        \label{eq:axi2} \end{eqnarray}
\end{subequations}
For the inclusion of the zero in the axial-vector meson form factor we also make 
here the changes
\[
\Omega^{(A)}_{i} \rightarrow \left(1-{\bf k}^2/U^2\right)\ \Omega^{(A)}_{i}\ ,
\]
with the same $U$-mass as used for the scalar mesons.
The motivation for the inclusion of a zero in the form factor here is again
 motivated by the quark-model, because for the axial-vector mesons one has the
configuration $Q\bar{Q}(^3P_1)$.                

 \begin{widetext}

As in Ref.~\cite{MRS89} in the derivation of the expressions for $\Omega_i^{(A)}$, 
given above, $M_{Y}$ and $M_{N}$ denote the mean hyperon and nucleon
mass, respectively \begin{math} M_{Y}=(M_{1}+M_{3})/2 \end{math}
and \begin{math} M_{N}=(M_{2}+M_{4})/2 \end{math},
 and $m$ denotes the mass of the exchanged meson.
Moreover, the approximation                            
       \begin{math}
              1/ M^{2}_{N}+1/ M^{2}_{Y}\approx
              2/ M_{N}M_{Y},
       \end{math}
is used, which is rather good since the mass differences
between the baryons are not large.

\subsection{One-Boson-Exchange Interactions in Configuration Space}
\label{sect.IIIb}
\begin{enumerate}
\item[a)] 
For $X=P$ the local configuration space potentials are given in
Ref.~\cite{MRS89}. Here, we give the non-local Graz- corrections. From the 
Fourier transform of the $\Omega^{(P)}_{2b,3b}$ contributions and (\ref{eq:psx2}) 
we have
\begin{eqnarray}
  \Delta V_{PS}(r) &=& \frac{f_{13}f_{24}}{4\pi}\cdot 
 \frac{m^3}{m_\pi^2}\cdot\left\{\vphantom{\frac{A}{A}}
 \frac{1}{3} (\mbox{\boldmath $\sigma$}_1\cdot\mbox{\boldmath $\sigma$}_2)
 \left(\nabla^2\phi_C^1+\phi_C^1\nabla^2\right) 
  \right.\nonumber\\ && \left. \vphantom{\frac{A}{A}}
 +\left(\nabla^2\phi_T^0\ S_{12} 
+ \phi_T^0 S_{12} \nabla^2\right)\right\}
 /(4M_YM_N)\ ,
 \label{eq:3.14}\end{eqnarray}
where $\phi_C^0, \phi_C^1, \phi_T^0$ are defined in \cite{NRS78,MRS89}, and are
functions of $(m,r,\Lambda)$.
 \item[b)] 
 Again, for $X=V,D$ we refer to the configuration space potentials 
in Ref.~\cite{MRS89}. For $X=S$ we give here the additional terms w.r.t. those 
in \cite{MRS89}, which are due to the zero in the scalar form factor. They are
\begin{eqnarray}
 &&  \Delta V_{S}(r) = - \frac{m}{4\pi}\ \frac{m^2}{U^2}\
 \left[ g^S_{13} g^S_{24}\left\{
 \left[\phi^1_C - \frac{m^2}{4M_YM_N} \phi^2_C\right]
+\frac{m^2}{2M_YM_N}\phi^1_{SO}\ {\bf L}\cdot{\bf S}
 \right.\right. \nonumber\\ && \left.\left.
+\frac{m^4}{16M_Y^2M_N^2}\phi^1_T\ Q_{12}  
 + \frac{m^2}{4M_YM_N}\frac{M_N^2-M_Y^2}{M_YM_N}\ \phi^{(1)}_{SO}\cdot
 \frac{1}{2}(\mbox{\boldmath $\sigma$}_1-\mbox{\boldmath $\sigma$}_2)\cdot{\bf L}
  \right\}\right]\ . 
 \label{eq:3.15}\end{eqnarray}


\item[c)] 
For the axial-vector mesons, the configuration space potential 
 corresponding to (\ref{eq:axi1}) is             
\begin{eqnarray}
 &&  V_{A}^{(1)}(r) = - \frac{g_{A}^{2}}{4\pi}\ m  \left[
 \phi^{0}_{C}\ (\mbox{\boldmath $\sigma$}_1\cdot\mbox{\boldmath $\sigma$}_2) 
  -\frac{1}{12M_YM_N}
  \left( \nabla^{2} \phi^{0}_{C}+\phi^{0}_{C}\nabla^{2}\right)
 (\mbox{\boldmath $\sigma$}_1\cdot\mbox{\boldmath $\sigma$}_2) 
 \right. \nonumber \\ && \nonumber \\ & & \left. \hspace*{1.4cm}
   + \frac{3m^{2}}{4M_YM_N}\ \phi^{0}_{T}\ S_{12}
 +\frac{m^{2}}{2M_YM_N}\ \phi^{0}_{SO}(m,r)\ {\bf L}\cdot{\bf S}
 \right. \nonumber \\ && \nonumber \\ & & \left. \hspace*{1.4cm}
 + \frac{m^2}{4M_YM_N}\frac{M_N^2-M_Y^2}{M_YM_N}\ \phi^{(0)}_{SO}\cdot
 \frac{1}{2}(\mbox{\boldmath $\sigma$}_1-\mbox{\boldmath $\sigma$}_2)\cdot{\bf L}
  \right]\ .
 \label{eq:3.16}\end{eqnarray}
The configuration space potential corresponding to (\ref{eq:axi2}) is             
\begin{eqnarray}
   V_{A}^{(2)}(r) &=&  \frac{g_{A}^{2}}{4\pi}\ m\  \left[
  \frac{1}{3}(\mbox{\boldmath $\sigma$}_1\cdot\mbox{\boldmath $\sigma$}_2) 
  \phi^{1}_{C} +\frac{1}{12M_YM_N} (
 (\mbox{\boldmath $\sigma$}_1\cdot\mbox{\boldmath $\sigma$}_2) 
  \left( \nabla^{2} \phi^{1}_{C}+ \phi^{1}_{C}\nabla^{2} \right) 
 \right. \nonumber\\ && \left.
  + S_{12}\ \phi^{0}_{T} 
  +\frac{1}{4M_YM_N} \left( \nabla^{2} \phi^{0}_{T} S_{12} +
  \phi^{0}_{T} S_{12} \nabla^{2} \right)
 \right]\ ,
\label{eq:3.17}\end{eqnarray}
The extra contribution to the potentials coming from the zero in the axial-vector
meson form factor are obtained from the expression (\ref{eq:3.16}) by making 
substitutions as follows
\begin{eqnarray}
   \Delta V_{A}^{(1)}(r) &=&  V_{A}^{(1)}\left(\phi_C^0 \rightarrow \phi_C^1,
 \phi_T^0 \rightarrow \phi_T^1, \phi_{SO}^0 \rightarrow \phi_{SO}^1\right)
 \cdot\frac{m^2}{U^2}\ .
\label{eq:3.17b}\end{eqnarray}
Note that we do not include the similar $\Delta V_A^{(2)}(r)$ since they involve
${\bf k}^4$-terms in momentum-space. 
\end{enumerate}

 \end{widetext}

\subsection{PS-PS-exchange Interactions in Configuration Space}                       
\label{sect.IIIc}
In Fig.~\ref{bwfig} and Fig.~\ref{tmofig} the included two-meson exchange 
graphs are shown schematically. 
The Bruckner-Watson (BW) graphs \cite{Bru53} contain in all three intermediate 
states both mesons and nucleons. The Taketani-Machida-Ohnuma (TMO) graphs 
\cite{Tak52} have one intermediate state with only nucleons.
Explicit expression for  
$K^{irr}(BW)$ and $K^{irr}(TMO)$ were derived \cite{Rij91}, where also the 
terminology BW and TMO is explained.
The TPS-potentials for nucleon-nucleon have been given in detail in \cite{RS96a}.
The generalization to baryon-baryon is similar to that for the {\rm OBE}-potentials.
So, we substitute $M \rightarrow \sqrt{M_YM_N}$, and include all PS-PS 
possibilities with coupling constants as in the {\rm OBE}-potentials. 
As compared to nucleon-nucleon in \cite{RS96a} here we have included in addition 
the potentials with double K-exchange.  The masses
are the physical pseudoscalar meson masses. For the intermediate two-baryon
states we take into account of the different thresholds.
We have not included uncorrelated PS-vector, PS-scalar, or PS-diffractive 
exchange. This because the range of these potentials is similar to 
those of the vector-, scalar-, and axial-vector-potentials. Moreover, for 
potentially large potentials, in particularly those with scalar mesons involved,
there will be very strong cancellations between the planar- and crossed-box
contributions.

\subsection{{\rm MPE}-exchange Interactions}
\label{sect.IIId}
In Fig.~\ref{pairfig} both the one-pair graphs and the two-pair graphs are shown.
In this work we include only the one-pair graphs. The argument for neglecting 
the two-pair graph is to avoid some 'double-counting'. Viewing the pair-vertex 
as containing heavy-meson exchange means that the contributions from $\rho(750)$
and $\epsilon=f_0(760)$ to the two-pair graphs is already accounted for by 
our treatment of the broad $\rho$ and $\epsilon$ {\rm OBE}-potential.
For a more complete discussion of the physics behind {\rm MPE} we refer to our 
previous papers \cite{Rij93,RS96b}.
The {\rm MPE}-potentials for nucleon-nucleon have been given in \cite{RS96b}.
The generalization to baryon-baryon is similar to that for the TPS-potentials.
For the intermediate two-baryon
states we neglect the different two-baryon thresholds. This because,
although in principle possible, it complicates the computation of the 
potentials considerably. 
The generalization of the pair-couplings to baryon-baryon is described in paper II
\cite{RY04b}, section III.
Also here in $N\!N$, we have in addition to  \cite{RS96b}  
included the pair-potentials with $K\otimes K$-, $K\otimes K^*$-, 
and $K\otimes \kappa$-exchange.
The convention for the {\rm MPE} coupling constants is the same as in \cite{RS96b}.

\subsection{The Schr\"{o}dinger equation with Non-local potential}
\label{sect.IIIe}
The non-local potentials are of the central-, spin-spin, and tensor-type. The method 
of solution of the Schr\"{o}dinger equation for nucleon-nucleon is described in 
\cite{NRS78} and \cite{Graz78}. Here, the non-local tensor is in momentum space 
of the form ${\bf q}^2\ \tilde{v}_T({\bf k})$. For a more general treatment of the
non-local potentials see \cite{Rij98}.

\section{ ESC-couplings and the {\rm QPC}-model}                     
\label{sec:4} 
 According to the Quark-Pair-Creation ({\rm QPC}) model, in the $^3P_0$-version
 \cite{Mic69}, the
baryon-baryon-meson couplings are given in terms of the quark-pair creation 
constant $\gamma_M$, and the radii of the (constituent) gaussian quark wave
functions, by \cite{Yao73,Yao75}
\begin{eqnarray*}
  g_{BBM}(\mp) &=& 2 \left(9\pi\right)^{1/4} \gamma_M\ 
  X_M\left(I_M,L_M,S_M,J_M\right)\ F^{(\mp)}_M\ ,
\end{eqnarray*}
where $X_M(\ldots)$ is a isospin, spin etc. recoupling coefficient, and 
\begin{eqnarray*}
 F^{(-)} &=& \left(m_MR_M\right)^{3/2}
 \left(\frac{3R_B^2}{3R_B^2+R_M^2}\right)^{3/2}
 \left(\frac{4R_B^2+R_M^2}{3R_B^2+R_M^2}\right)\ ,
 \nonumber\\ 
 F^{(+)} &=& \left(m_MR_M\right)^{1/2}
 \left(\frac{3R_B^2}{3R_B^2+R_M^2}\right)^{3/2}
 \frac{4R_M^2}{(3R_B^2+R_M^2)}   
\end{eqnarray*}
are coming from the overlap integrals.
Here, the superscripts $\mp$ refer to the parity of the mesons $M$: $(-)$ for 
$J^{PC}=0^{+-}, 1^{--}$, and $(+)$ for $J^{PC}= 0^{++}, 1^{++}$. The radii of the 
baryons, in this case nucleons, and the mesons are respectively denoted by $R_B$ and $R_M$.
 
The {\rm QPC}($^3P_0$)-model gives several interesting relations, such as  
\begin{equation}
\begin{array}{cccccccc}
 g_\omega &=& 3 g_\rho &,& g_\epsilon &=& 3 g_{a_0} & , \\             
 g_{a_0} &\approx & g_\rho &,&  g_\epsilon & \approx & g_\omega & .            
\end{array}
\end{equation}
We see here an interesting link between the vector-meson and the scalar-meson couplings,
which is not totally surprising, because the scalar polarization-vector $\epsilon_0$
of the vector mesons 
in the quark-model is realized by a $Q\bar{Q}(^3P_0)$-state. This is the same
state as for the scalar mesons in the $Q\bar{Q}$-picture.

From $\rho \rightarrow e^+ e^-$, employing the current-field-identities (C.F.I's)
one can derive, see for example \cite{Roy67}, the following relation with the {\rm QPC}-model      
\begin{equation}
 f_\rho = \frac{m_\rho^{3/2}}{\sqrt{2}|\psi_\rho(0)|} \Leftrightarrow
 \gamma_M\left(\frac{2}{3\pi}\right)^{1/2}
 \frac{m_\rho^{3/2}}{|'\psi_\rho(0)'|}\ ,           
\label{eq:gam0}\end{equation}
which, neglecting the difference between the wave functions on the left and
right hand side, gives for the pair creation constant 
$\gamma_M \rightarrow \gamma_0=\frac{1}{2}\sqrt{3\pi} =1.535$. However, 
since in the {\rm QPC}-model gaussian wave functions are used, the $Q\bar{Q}$-potential
is a harmonic-oscillator one. This does not account for the $1/r$-behavior,
due to one-gluon-exchange (OGE), at short distance. This implies a OG-correction
\cite{LP96} to the wave function, which gives for $\gamma_M$ \cite{Chai80}
\begin{equation}
 \gamma_M = \gamma_0 \left(1-\frac{16}{3}\frac{\alpha(m_M)}{\pi}\right)^{-1/2}\ .
\label{eq:gam}\end{equation}
In Table~\ref{tab.gam} $\gamma_M(\mu)$ is shown,
using from \cite{PDG02} the parameterization
\begin{equation}
 \alpha_s(\mu) = 4\pi/\left(\beta_0\ln(\mu^2/\Lambda_{QCD}^2)\right)\ ,
\label{eq:alphas}\end{equation}
with $\Lambda_{QCD} = 100$ MeV and $\beta_0 = 11-\frac{2}{3} n_f$ for $n_f=3$.
\begin{table}[hbt]
 \caption{Pair-creation constant $\gamma_M$ as function of $\alpha_s$. }       
\begin{center}
\begin{ruledtabular}
\begin{tabular}{ccc} & \\
  $\mu$ [GeV] & $\alpha_s(\mu)$   & $\gamma_M(\mu)$     \\
        &        &               \\
\hline
        &        &               \\
  $\infty$ &  0.00         &  1.535        \\
  80.0  &  0.10         &  1.685        \\
  35.0  &  0.20         &  1.889        \\
  1.05  &  0.30         &  2.191        \\
  0.55  &  0.40         &  2.710        \\
  0.40  &  0.50         &  3.94         \\
  0.35  &  0.55         &  5.96         \\
        &        &               \\
\end{tabular}
\end{ruledtabular}
\end{center}
\label{tab.gam}
\end{table}
From this table one sees that at the scale of $m_M \approx 1$ GeV a value 
$\gamma_M = 2.19$ is reasonable. 
This value we will use later
when comparing the {\rm QPC}-model predictions and the ESC04-model
coupling constants.
As remarked in \cite{Chai80} the correction to $\gamma_0$ is not small, and 
therefore should be seen as an indication.\\
In Table~\ref{tab.3p0} we show the $^3P_0$-model results and the values obtained 
in the ESC04-fit. 
In this table we fixed $\gamma_M = 2.19$ for the vector-, scalar-, and 
axial-vector-mesons, for $R_B = 0.54$ fm. This 'effective' radius is choosen
from \cite{Yao73}, where it was determined using the Regge slopes.
Here, one has to realize that the {\rm QPC}-predictions are kind of "bare" couplings,
which allows vertex corrections from meson-exchange.
For the pseudoscalar, a different value has to be used, showing indeed 
some 'running'-behavior as expected from QCD. 
In \cite{Chai80}, for the decays $\rho, \epsilon \rightarrow 2\pi$ etc. it was found
$\gamma_\pi =3.33$, whereas we need here $\gamma_\pi=4.84$. 
Of course, there are several ways to change this by, for example, using other 
'effective' meson-radii.
For the mesonic decays of the charmonium states $\gamma_\psi=1.12$.
One notices the similarity between the {\rm QPC}($^3P_0$)-model predictions 
and the fitted couplings.


\begin{table}[hbt]
 \caption{ESC04 Couplings and $^3P_0$-Model Relations.}       
\begin{ruledtabular}
\begin{tabular}{lccccc} & & & & &  \\
  Meson          & $r_M[{\it fm}]$ & $X_M$ & $\gamma_M$ & $^3P_0$ & ESC04 \\
                 &       &       &          &         &      \\
\hline
                 &       &       &          &         &      \\
 $\pi(140)$      & 0.66  & $5/6$ &  4.84    & $f=0.26$& $f=0.26$ \\
                 &       &       &          &         &      \\
 $\rho(770)$     & 0.66  & $1  $ &  2.19    & $g=0.93$& $g=0.78$ \\
                 &       &       &          &         &      \\
 $\omega(783)$   & 0.66  & $3  $ &  2.19    & $g=2.86$& $g=3.12$ \\
                 &       &       &          &         &      \\
 $a_0(962)$      & 0.66  & $1 $  &  2.19    & $g=0.93$& $g=0.81$ \\
                 &       &       &          &         &      \\
 $\epsilon(760)$ & 0.66  &$3 $   &  2.19    & $g=2.47$& $g=2.87$ \\
                 &       &       &          &         &      \\
 $a_1(1270)$     & 0.66  &$5\sqrt{2}/6$&  2.19    & $g=2.51$& $g=2.42$ \\
                 &       &       &          &         &      \\
\end{tabular}
\end{ruledtabular}
\label{tab.3p0}         
\end{table}

Finally, we notice that the Schwinger relation \cite{Sch68}
\begin{eqnarray}
 g_{NNa_1} &\approx& \frac{m_{a_1}}{m_\pi} f_{NN\pi}\ ,                  
\end{eqnarray}
is also rather well satisfied, both in the {\rm QPC}-model and the ESC04-fit.

\section{ ESC-model , Results}                                  
\label{sec:7} 
\subsection{ Parameters and Nucleon-nucleon Fit}                                  
\label{sec:7a} 
During the searches fitting the {\it NN}-data with the present ESC-model ESC04,  
it was found that the {\rm OBE}-couplings could be constraint successfully
using the 'naive' {\rm QPC}-predictions as a guidance \cite{Mic69}. Although these 
predictions, see section \ref{sec:4}, are 'bare' ones, we kept during the 
searches all {\rm OBE}-couplings rather closely in the neighborhood of these predictions.
Also, it appeared that we could either fix all $F/(F+D)$-ratios 
to those as suggested by the {\rm QPC}-model, 
or apply the same strategy as for the {\rm OBE}-couplings.                      

The meson nonets contain {rm SU(3)} octet and mixed octet-singlet members.
We assign in principle cut-offs $\Lambda_8$ and $\Lambda_1$ to the octets and
singlets respectively. However, because of the octet-singlet mixings for the 
$I=0$ members, and the use of the physical mesons in the potentials, we use 
$\Lambda_1$ for all $I=0$-mesons. We have as free cut-off parameters 
$(\Lambda^P_8, \Lambda^V_8, \Lambda^S_8)$, and similarly a set for the singlets.
For the axial-vector mesons we use a single cut-off $\Lambda^A$.


The treatment of the broad mesons $\rho$ and $\epsilon$ is the same as in the 
{\rm OBE}-models \cite{NRS78,MRS89}. 
In this treatment a broad meson is approximated by two narrow mesons. The mass and
width of the broad meson determines the masses $m_{1,2}$ and the weights $\beta_{1,2}$ 
of these narrow ones. For the $\rho$-meson the same parameters are used as in
\cite{NRS78,MRS89}. However, for $\epsilon= f_0(760)$, assuming \cite{NRS78}
$m_\epsilon=760$ MeV and $\Gamma_\epsilon = 640$ MeV, the Bryan-Gersten parameters
\cite{Bry72} are used: $ m_1=496.39796$ MeV, $m_2=1365.59411$ MeV, and 
$\beta_1=0.21781, \beta_2=0.78219$.

The 'mass' of the diffractive exchanges were all fixed to $m_P=309.1$ MeV.

Summarizing the parameters we have for {\it NN}:
\begin{enumerate}
\item {\rm QPC}-constrained: $f_{NN\pi},f_{NN\eta'}, 
 g_{NN\rho}, g_{NN\omega}$, \\
 $f_{NN\rho},f_{NN\omega}, g_{NNa_1}, g_{a_0},g_{NN\epsilon}, g_{NNA_2},g_{NNP}$, 
\item Pair couplings: $g_{NN(\pi\pi)_1},f_{NN(\pi\pi)_1}, g_{NN(\pi\rho)_1}$,\\
 $g_{NN\pi\omega}, g_{NN\pi\eta}, g_{NN\pi\epsilon}$, 
\item Cut-off masses: $\Lambda_8^P, \Lambda_8^V,  \Lambda_8^S,
\Lambda_1^V,  \Lambda_1^S, \Lambda^A$.
\end{enumerate}
The pair coupling $g_{NN(\pi\pi)_0}$ was kept fixed at a small, 
but otherwise arbitrary value.

Together with the fit to the 1993 Nijmegen representation of the $\chi^2$-hypersurface of the 
{\it NN}-scattering data below $T_{lab}=350$ MeV \cite{Sto93}, also some  
low-energy parameters were fitted: the {\it np} and {\it nn} scattering lengths and effective ranges for 
the $^1S_0$, and the binding energy of the deuteron $E_B$.

\begin{figure}   
\resizebox{8.cm}{11.43cm}        
 {\includegraphics[50,50][554,770]{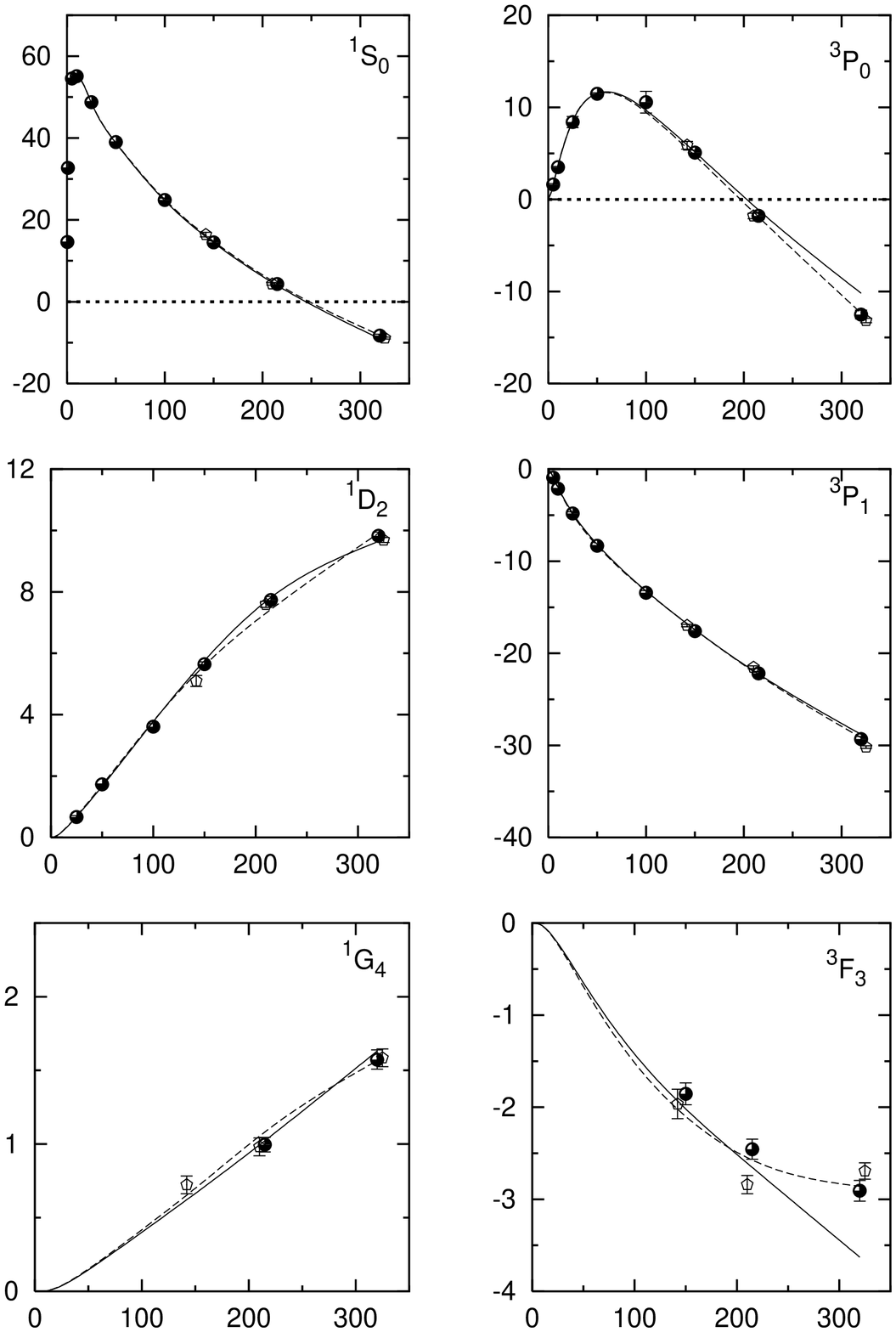}}
\caption{Solid line: proton-proton $I=1$ phase shifts (degrees), as a function of
 $T_{\rm lab}$(MeV), for the ESC04-model. 
 The dashed line: the m.e. phases of the Nijmegen93 PW-analysis \cite{Sto93}. 
 The black dots:  the s.e. phases of the Nijmegen93 PW-analysis.
 The diamonds:  Bugg s.e. \cite{Bugg92}.}
\label{ppi1.fig}
 \end{figure}

\begin{figure}   
\resizebox{8.cm}{11.43cm}
 {\includegraphics[50,50][554,770]{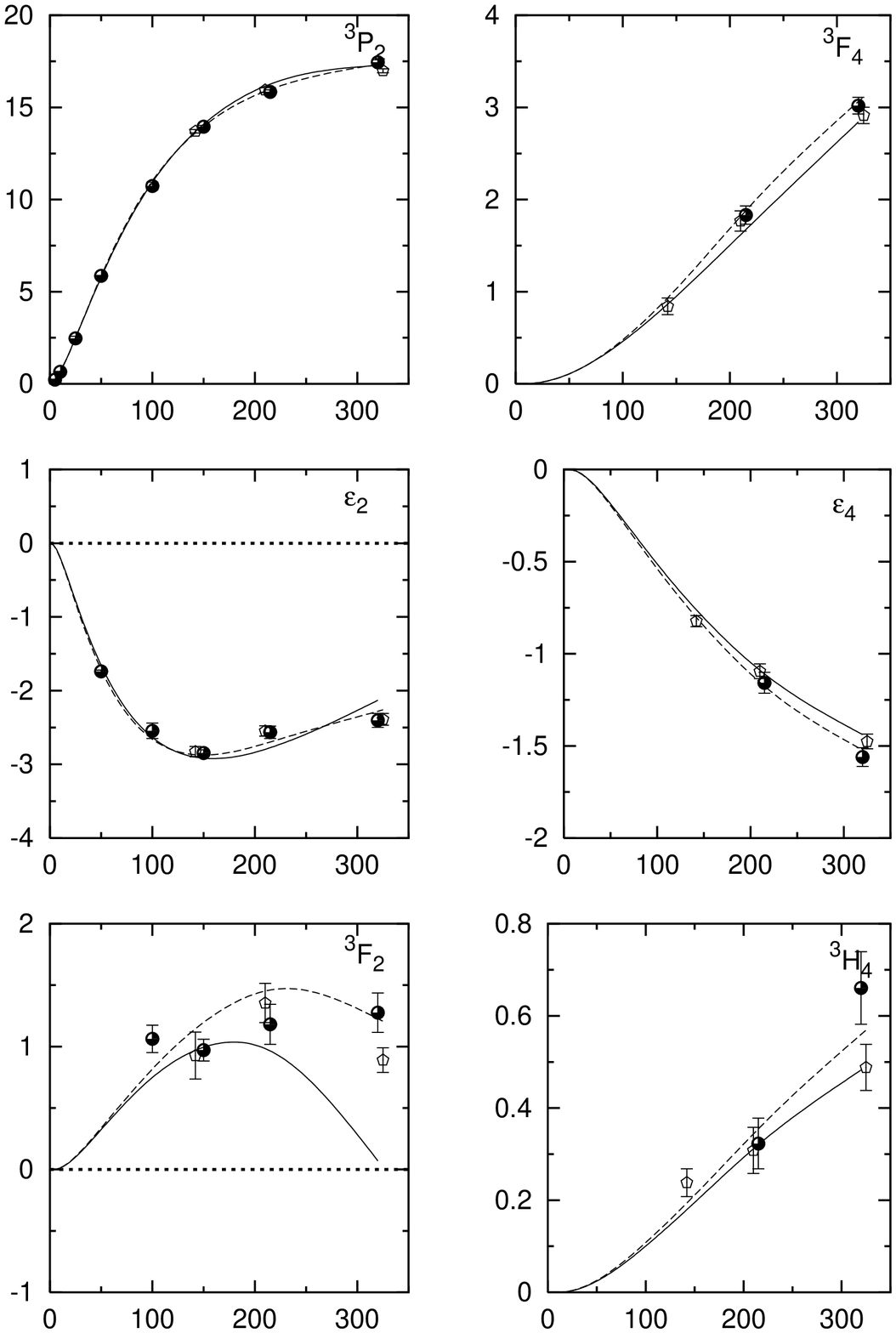}}
\caption{Solid line: proton-proton $I=1$ phase shifts (degrees), as a function of
 $T_{\rm lab}$(MeV), for the ESC04-model. 
 The dashed line: the m.e. phases of the Nijmegen93 PW-analysis \cite{Sto93}. 
 The black dots:  the s.e. phases of the Nijmegen93 PW-analysis.
 The diamonds:  Bugg s.e. \cite{Bugg92}.}
\label{ppi1c.fig}
 \end{figure}

\begin{figure}   
\resizebox{8.cm}{7.7cm}
 {\includegraphics[50,280][554,770]{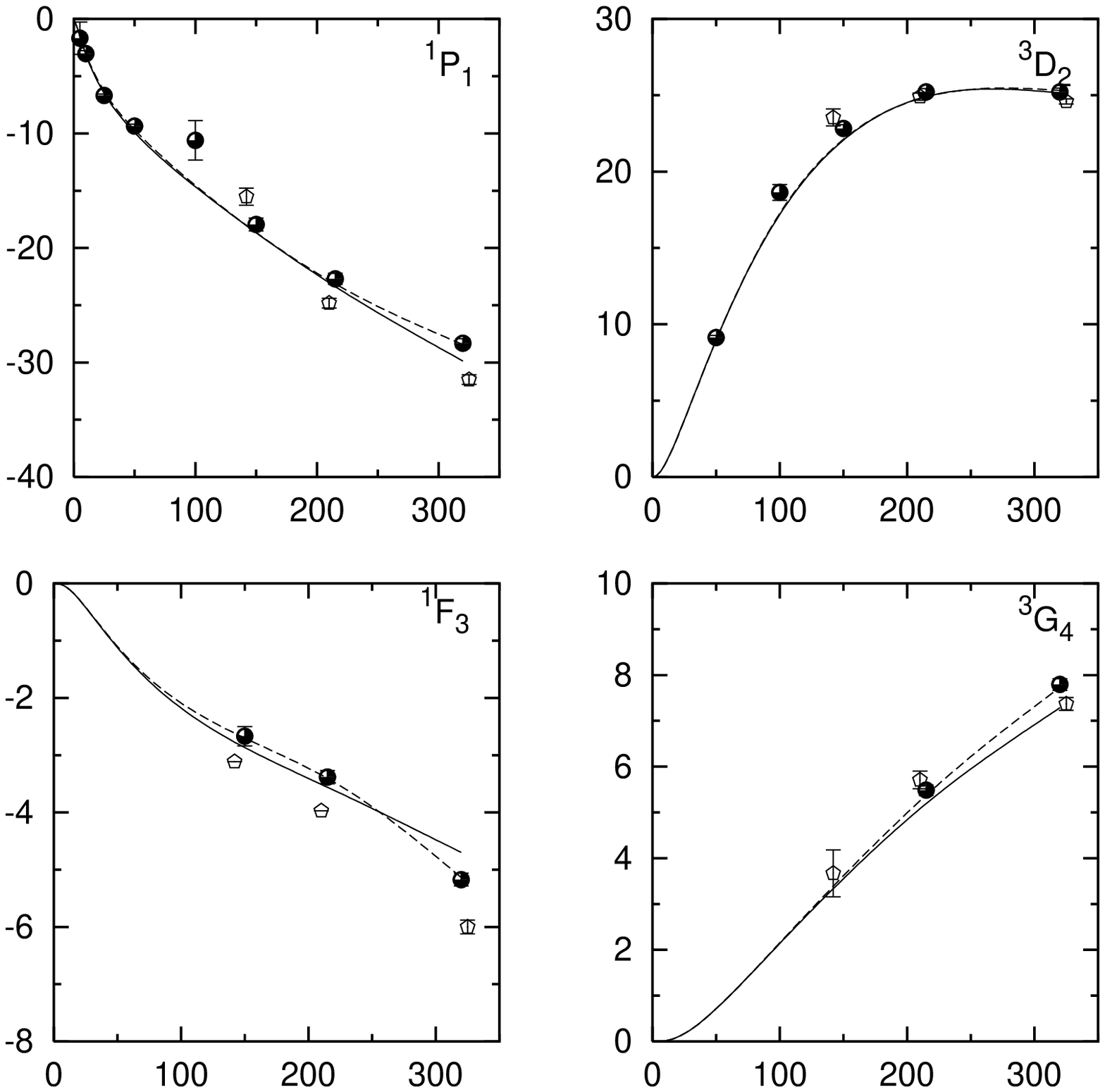}}
\caption{Solid line: neutron-proton $I=0$ phase shifts (degrees), as a function of
 $T_{\rm lab}$(MeV) for the ESC04-model. 
 The dashed line: the m.e. phases of the Nijmegen93 PW-analysis \cite{Sto93}. 
 The black dots:  the s.e. phases of the Nijmegen93 PW-analysis.
 The diamonds:  Bugg s.e. \cite{Bugg92}.}
\label{npi0.fig}
 \end{figure}

\begin{figure}   
\resizebox{8.cm}{11.43cm}
 {\includegraphics[50,50][554,770]{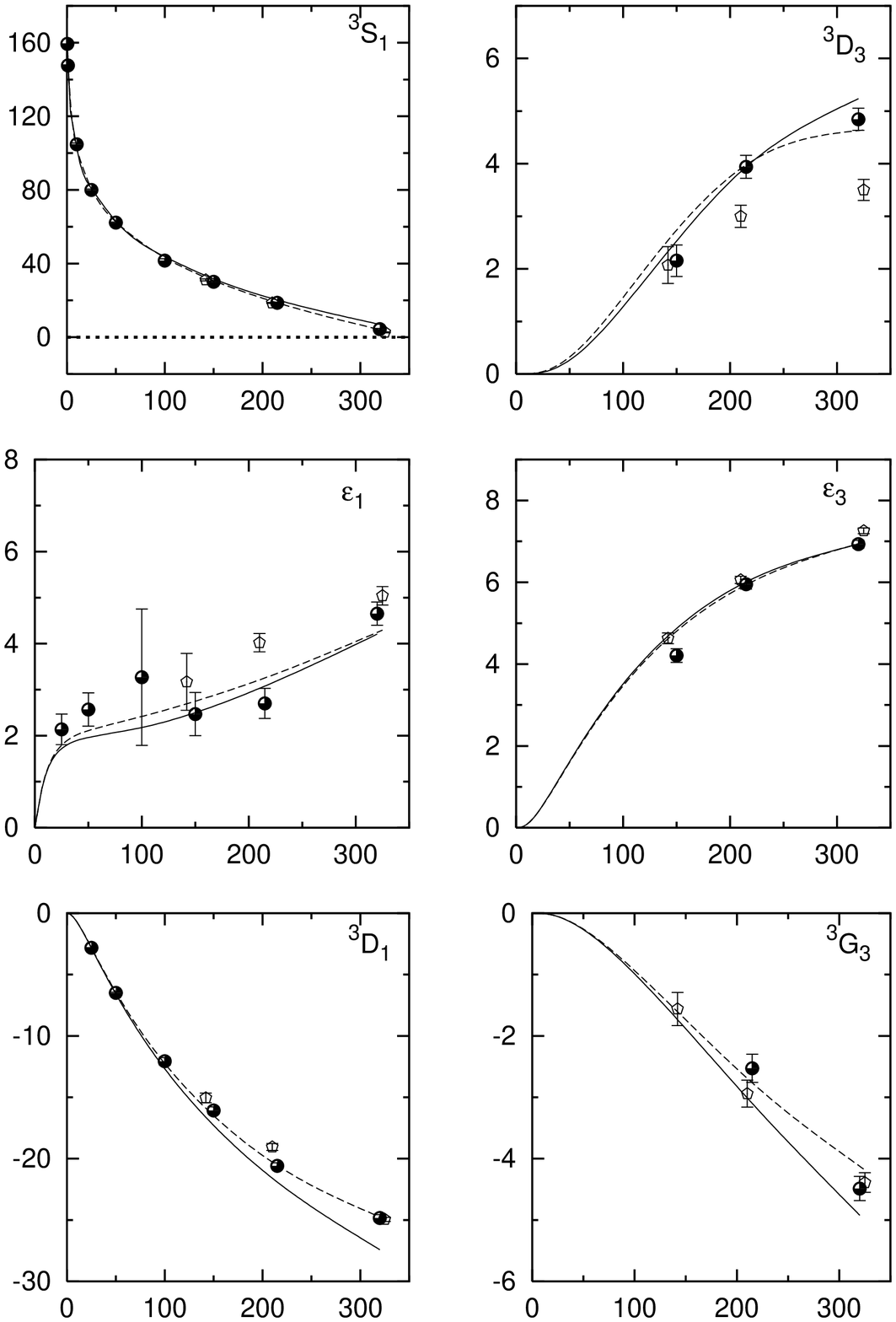}}
\caption{Solid line: neutron-proton $I=0$ phase shifts (degrees), as a function of
 $T_{\rm lab}$(MeV) for the ESC04-model. 
 The dashed line: the m.e. phases of the Nijmegen93 PW-analysis \cite{Sto93}. 
 The black dots:  the s.e. phases of the Nijmegen93 PW-analysis.
 The diamonds:  Bugg s.e. \cite{Bugg92}.}
\label{npi0c.fig}
 \end{figure}

We obtained for the phase shifts a $\chi^2/N_{\rm data} =1.155$. 
The phase shifts are shown in
Table's \ref{tab.nnphas1} and \ref{tab.nnphas2}, 
and also in Fig.'s~\ref{ppi1.fig}-\ref{npi0c.fig}.  
In Table~\ref{tab.chidistr} the distribution of the $\chi^2$ for ESC04 is 
shown for the ten energy bins used in the single-energy (s.e.) phase shift analysis,
and compared with that of the updated partial-wave analysis \cite{Klo93}.

\begin{table}[h]
\caption{ ESC04 {\it pp} and {\it np} nuclear-bar phase shifts in degrees.}
\begin{ruledtabular}
\begin{tabular}{crrrrr}  & & & & & \\
 $T_{\rm lab}$ & 0.38& 1 & 5  & 10 & 25 \\ \hline
     &    &     &     &     &     \\
 $\sharp$ {\rm data} &144  & 68  & 103 & 290& 352 \\
     &    &     &     &     &   \\
$\Delta \chi^{2}$& 20  & 38  & 17  & 34  & 12  \\
     &    &     &     &     &    \\ \hline
     &    &     &     &     &    \\
 $^{1}S_{0}(np)$ & 54.58  & 61.89 & 63.04 & 59.13& 49.66  \\
 $^{1}S_{0}$ & 14.62  & 32.63 & 54.76 & 55.16& 48.58  \\
 $^{3}S_{1}$ & 159.38 & 147.76& 118.21& 102.66& 80.76\\
 $\epsilon_{1}$ & 0.03  & 0.11 & 0.67 & 1.14 & 1.72 \\
 $^{3}P_{0}$ & 0.02   &  0.13 & 1.55 & 3.67 & 8.50  \\
 $^{3}P_{1}$ & -0.01  &-0.08  &-0.87  & -1.98  & -4.78  \\
 $^{1}P_{1}$ & -0.05  &-0.19  &-1.52  & -3.12  & -6.49  \\
 $^{3}P_{2}$ &  0.00  & 0.01  & 0.22  &  0.66  &  2.49  \\
 $\epsilon_{2}$ &-0.00  &-0.00 &-0.05 &-0.19 &-0.78 \\
 $^{3}D_{1}$ & 0.00  &-0.01  &-0.19  & -0.69  & -2.85  \\
 $^{3}D_{2}$ & 0.00  & 0.01  & 0.22  &  0.86  &  3.73  \\
 $^{1}D_{2}$ & 0.00  & 0.00  & 0.04  &  0.16  &  0.68  \\
 $^{3}D_{3}$ & 0.00  & 0.00  & 0.00  &  0.00  &  0.04  \\
 $\epsilon_{3}$ & 0.00  & 0.00 & 0.01 & 0.08 & 0.56 \\
 $^{3}F_{2}$ & 0.00  & 0.00  & 0.00  &  0.01  &  0.10  \\
 $^{3}F_{3}$ & 0.00  & 0.00  &-0.00  & -0.03  & -0.22  \\
 $^{1}F_{3}$ & 0.00  & 0.00  &-0.01  & -0.07  & -0.42  \\
 $^{3}F_{4}$ & 0.00  & 0.00  & 0.00  &  0.00  &  0.02  \\
 $\epsilon_{4}$ & 0.00  & 0.00 & 0.00 &-0.00 &-0.05 \\
     &    &     &     &     &    \\
\end{tabular}
\end{ruledtabular}
\label{tab.nnphas1}   
\end{table}

\begin{table}[h]
\caption{ ESC04 {\it pp} and {\it np} nuclear-bar phase shifts in degrees.}
\begin{ruledtabular}
\begin{tabular}{crrrrr}  & & & & & \\
 $T_{\rm lab}$ &  50 &100 & 150 & 215 & 320 \\ \hline
     &    &     &     &     &     \\
 $\sharp$ data &572  &399  &676  & 756& 954 \\
     &    &     &     &     &   \\
$\Delta \chi^{2}$& 118 & 29  & 114 & 137 & 337 \\
     &    &     &     &     &    \\ \hline
     &    &     &     &     &    \\
 $^{1}S_{0}(np)$ & 38.81  & 24.24 & 13.80 & 3.27 & -9.80  \\
 $^{1}S_{0}$ & 38.77  & 24.71 & 14.42 & 3.97 & -9.05  \\
 $^{3}S_{1}$ & 63.03  & 43.79 & 31.66 & 20.27 & 6.93 \\
 $\epsilon_{1}$ & 1.96  & 2.18 & 2.50 & 3.08 & 4.21 \\
 $^{3}P_{0}$ & 11.51  &   9.68& 5.14 &-1.13 &-10.19 \\
 $^{3}P_{1}$ & -8.16  &-13.22 &-17.43 & -22.24 & -28.81 \\
 $^{1}P_{1}$ & -9.92  &-14.65 &-18.67 & -22.37 & -29.87 \\
 $^{3}P_{2}$ &  5.78  & 10.94 & 14.09 &  16.26 &  17.28 \\
 $\epsilon_{2}$ &-1.66  &-2.63 &-2.92 &-2.77 &-2.13 \\
 $^{3}D_{1}$ &-6.58  &-12.67 &-17.29 & -21.90 & -27.41 \\
 $^{3}D_{2}$ & 8.97  & 17.20 & 22.06 &  24.92 &  25.15 \\
 $^{1}D_{2}$ & 1.67  & 3.77  & 5.76  &  7.82  &  9.65  \\
 $^{3}D_{3}$ & 0.27  & 1.28  & 2.53  &  3.94  &  5.24  \\
 $\epsilon_{3}$ & 1.62  & 3.52 & 4.87 & 6.01 & 6.93 \\
 $^{3}F_{2}$ & 0.32  & 0.75  & 1.00  &  0.97  &  0.07  \\
 $^{3}F_{3}$ &-0.65  &-1.42  &-2.02  & -2.65  & -3.63  \\
 $^{1}F_{3}$ &-1.12  &-2.18  &-2.87  & -3.56  & -4.70  \\
 $^{3}F_{4}$ & 0.11  & 0.46  & 0.95  &  1.67  &  2.84  \\
 $\epsilon_{4}$ &-0.19  &-0.51 &-0.81 &-1.11 &-1.44 \\
 $^{3}G_{3}$ &-0.27 &-0.99  &-1.88  &-3.10  & -4.92  \\
 $^{3}G_{4}$ & 0.72  & 2.14  & 3.56  &  5.20  &  7.29  \\
 $^{1}G_{4}$ & 0.15  & 0.40  & 0.67  &  1.02  &  1.63  \\
 $^{3}G_{5}$ &-0.05  &-0.19  &-0.32  & -0.42  & -0.43  \\
 $\epsilon_{5}$ & 0.21  & 0.72 & 1.26 & 1.90 & 2.75 \\
     &    &     &     &     &    \\
\end{tabular}
\end{ruledtabular}
\label{tab.nnphas2}   
\end{table}
 
\begin{table}
\caption{ESC04 Low energy parameters: S-wave scattering lengths and 
effective ranges, deuteron binding energy $E_B$, and electric 
quadrupole $Q_e$.}
\begin{ruledtabular}
\begin{tabular}{ccccc} & & & & \\
     & \multicolumn{3}{c}{experimental data}& ESC04 
     \\ &&&& \\ \hline
 $a_{pp}(^1S_0)$ & -7.823 & $\pm$ & 0.010 & -7.770        \\
 $r_{pp}(^1S_0)$ &  2.794 & $\pm$ & 0.015 &  2.753        \\ \hline
 $a_{np}(^1S_0)$ & -23.715 & $\pm$ & 0.015 & -23.860 \\
 $r_{np}(^1S_0)$ &  2.760 & $\pm$ & 0.030 &  2.787 \\ \hline
 $a_{nn}(^1S_0)$ & -18.70     & $\pm$ & 0.60  & -18.63 \\
 $r_{nn}(^1S_0)$ &  2.75    & $\pm$ & 0.11  &  2.81\\ \hline
 $a_{np}(^3S_1)$ &  5.423 & $\pm$ & 0.005 &  5.404        \\
 $r_{np}(^3S_1)$ &  1.761 & $\pm$ & 0.005 &  1.749       \\ \hline
  $E_B$         &  -2.224644 & $\pm$ & 0.000046 & -2.224933 \\
  $Q_e$         &  0.286 & $\pm$ & 0.002 &  0.271        \\ 
\end{tabular}
\end{ruledtabular}
\label{tab.lowenergy}
\end{table}

\begin{table}[h]
\caption{Meson parameters employed in the potentials shown in
         Figs.~\protect\ref{obefig} to \protect\ref{pairfig}.
         Coupling constants are at ${\bf k}^{2}=0$.
         An asterisk denotes that the coupling constant is not searched,
         but constrained via {\rm SU(3)} or simply put to some value used in     
         previous work.
         The used widths of the $\rho$ and $\varepsilon$ are 146 MeV 
         and 640 MeV respectively.}
\begin{ruledtabular}
\begin{tabular}{crrrr}
meson & mass (MeV) & $g/\sqrt{4\pi}$ & $f/\sqrt{4\pi}$ & $\Lambda$ (MeV) \\
\hline
 $\pi$         &  138.04 \hspace{3.5mm} &           &   0.2621   &    829.90    \\
 $\eta$        &  548.80 \hspace{3.5mm} &           &   0.1673$^{\ast}$\hspace{-1.5mm}   & 900.00 \\
 $\eta'$       &  957.50 \hspace{3.5mm} &           &   0.1802   &    900.00    \\
 $\rho$        &  770.00 \hspace{3.5mm} &  0.7794   &   3.3166   &    782.38    \\
 $\omega$      &  783.90 \hspace{3.5mm} &  3.1242   &   0.0712   &    890.23    \\
 $\phi$        & 1019.50 \hspace{3.5mm} &--0.6957   &   1.2686$^{\ast}$\hspace{-1.5mm}   & 890.23 \\
 $a_1 $        & 1270.00 \hspace{3.5mm} &  2.4230   &            &    968.23    \\
 $f_1 $        & 1420.00 \hspace{3.5mm} &  1.4708   &            &  968.23      \\
 $f_1'$        & 1285.00 \hspace{3.5mm} &  0.5981$^{\ast}$\hspace{-1.5mm}   &            & 968.23 \\
 $a_{0}$       &  962.00 \hspace{3.5mm} &  0.8111   &            &   1161.27    \\
 $\varepsilon$ &  760.00 \hspace{3.5mm} &  2.8730   &            &   1101.62    \\
 $f_{0}$       &  993.00 \hspace{3.5mm} &--0.9669   &            &   1101.62    \\
 $a_{2}$       &  309.10 \hspace{3.5mm} &  0.0000   &            &              \\
 Pomeron       &  309.10 \hspace{3.5mm} &  2.2031   &            &              \\ 
\end{tabular}
\end{ruledtabular}
\label{tab.gobe}   
\end{table}

\begin{table}[h]
\caption{Pair-meson coupling constants employed in the ESC04 {\rm MPE}-potentials.     
         Coupling constants are at ${\bf k}^{2}=0$.}
\begin{ruledtabular}
\begin{tabular}{cclcc}
 $J^{PC}$ & {\rm SU(3)}-irrep & $(\alpha\beta)$  & $g/4\pi$  & $f/4\pi$ \\
\colrule
 $0^{++}$ & $\{1\}$  & $(\pi\pi)_{0}$   &  0.0000 &         \\
 $0^{++}$ & ,,       & $(\sigma\sigma)$ &  ---    &         \\
 $0^{++}$ &$\{8\}_s$ & $(\pi\eta)$      &--0.440  &         \\
 $0^{++}$ &          & $(\pi\eta')$     &  ---    &         \\
 $1^{--}$ &$\{8\}_a$ & $(\pi\pi)_{1}$   &  0.000  &  0.119  \\
 $1^{++}$ & ,,       & $(\pi\rho)_{1}$  &  0.835  &          \\
 $1^{++}$ & ,,       & $(\pi\sigma)$    &  0.022  &          \\
 $1^{++}$ & ,,       & $(\pi P)$        &  0.0    &          \\
 $1^{+-}$ &$\{8\}_s$ & $(\pi\omega)$    &--0.170  &          \\
\end{tabular}
\end{ruledtabular}
\label{tab.gpair}
\end{table}

\begin{table}
\caption{$\chi^2$ and $\chi^2$ per datum at the ten energy bins for the    
 Nijmegen93 partial-wave-analysis. $N_{\rm data}$ lists the number of data
 within each energy bin. The bottom line gives the results for the 
 total $0-350$ MeV interval.
 The $\chi^{2}$-excess for the ESC model is denoted    
 by  $\Delta\chi^{2}$ and $\Delta\hat{\chi}^{2}$, respectively.}  
\begin{ruledtabular}
\begin{tabular}{crrrrrr} & & & & & \\
 $T_{\rm lab}$ & $\sharp$ data & $\chi_{0}^{2}$\hspace*{5.5mm}&
 $\Delta\chi^{2}$&$\hat{\chi}_{0}^{2}$\hspace*{3mm}&
 $\Delta\hat{\chi}^{2}$ \\ &&&&& \\ \hline
0.383 & 144 & 137.5549 & 20.7 & 0.960 & 0.144  \\
  1   &  68 &  38.0187 & 52.4 & 0.560 & 0.771  \\
  5   & 103 &  82.2257 & 10.0 & 0.800 & 0.098  \\
  10  & 209 & 257.9946 & 27.5 & 1.234 & 0.095  \\
  25  & 352 & 272.1971 & 29.2 & 0.773 & 0.083  \\
  50  & 572 & 547.6727 &141.1 & 0.957 & 0.247  \\
  100 & 399 & 382.4493 & 32.4 & 0.959 & 0.081  \\
  150 & 676 & 673.0548 & 85.5 & 0.996 & 0.127  \\
  215 & 756 & 754.5248 &154.6 & 0.998 & 0.204  \\
  320 & 954 & 945.3772 &350.5 & 0.991 & 0.367  \\ \hline
      &    &     &     &     &    \\
Total &4233&4091.122& 903.9 &0.948 &0.208  \\
      &    &     &     &     &     \\
\end{tabular}
\end{ruledtabular}
\label{tab.chidistr} 
\end{table}

We emphasize that we use the single-energy (s.e.) phases and $\chi^2$-surfaces
\cite{Klo93}
only as a means to fit the {\it NN}-data. As stressed in \cite{Sto93} the 
Nijmegen s.e. phases have not much significance. The significant phases 
are the multi-energy (m.e.) ones, see the dashed lines in the figures.
One notices that the central value of the s.e. phases do not correspond
to the m.e. phases in general,
illustrating that there has been a certain amount
of noise fitting in the s.e. PW-analysis, see e.g. $\epsilon_1$ and $^1P_1$ 
at $T_{lab}=100$ MeV.
The m.e. PW-analysis reaches $\chi^2/N_{\rm data}=0.99$, using 
 39 phenomenological parameters plus normalization parameters, 
in total more than 50 free parameters.
The related phenomenological PW-potentials NijmI,II and Reid93 \cite{SKTS94},
with respectively 41, 47, and 50 parameters, all with $\chi^2/N_{\rm data}=1.03$.
This should be compared to the ESC-model, which has $\chi^2/N_{\rm data}=1.155$
using 20 parameters. These are 11 {\rm QPC}-constrained meson-nucleon-nucleon couplings,
6 meson-pair-nucleon-nucleon couplings, and 3 gaussian cut-off parameters.
From the figures it is obvious that the ESC-model deviates from the m.e.
PW-analysis at the highest energy for some partial waves. 
If we evaluate the $\chi^2$
for the first 9 energies only, we obtain $\chi^2/N_{\rm data} = 1.10$. 

In Table~\ref{tab.lowenergy} the results for the low energy parameters are given.
In order to discriminate between the $^1S_0$-wave for 
{\it pp}, {\it np}, and {\it nn}, we introduced 
some charge independence breaking by taking $g_{pp\rho} \neq g_{np\rho} \neq g_{nn\rho}$.
With this device we fitted the difference between the $^1S_0({\it pp})$ and $^1S_0({\it np})$ 
phases, and the different scattering lengths and effective ranges as well. We found
$g_{np\rho} = 0.71,\ g_{nn\rho} = 0.74$, which are not far from
$g_{pp\rho} = 0.78$, see Table~\ref{tab.gobe}.  

For $a_{nn}(^1S_0)$ we have used in the fitting the value from an investigation
of the {\it n}-{\it p} and {\it n}-{\it n} final state interaction 
in the $^2H({\it n},{\it nnp})$ reaction at 13 MeV
\cite{Gon99}. The value for $a_{\it nn}(^1S_0)$ is still somewhat in discussion.
Another recent determination \cite{Huh00} obtained e.g. $a_{\it nn}(^1S_0)= -16.27 \pm 0.40$
fm. Fitting with the latter value yields for the ESC04-model the value $-16.74$ fm.
Then, the quality of the fit to the phase shift analysis is the same, with small
changes to the parameters and phase shifts.
For a discussion of the theoretical and experimental situation w.r.t. these low 
energy parameters, see also \cite{Mil90}.

\subsection{ Coupling Constants }                                  
\label{sec:7b} 
In Table~\ref{tab.gobe} we show the {\rm OBE}-coupling constants and the 
 gaussian cut-off's $\Lambda$. The used  $\alpha =: F/(F+D)$-ratio's 
for the {\rm OBE}-couplings are:
pseudoscalar mesons $\alpha_{pv}=0.388$, 
vector mesons $\alpha_V^e=1.0, \alpha_V^m=0.387$, 
and scalar-mesons $\alpha_S=0.852$, which is computed using the physical 
$S^* = f_0(993)$ coupling etc..
In Table~\ref{tab.gpair} we show the {\rm MPE}-coupling constants.        
The used  $\alpha =: F/(F+D)$-ratio's for the {\rm MPE}-couplings are:
$(\pi\eta)$ etc. and $(\pi\omega)$ pairs $\alpha(\{8_s\})=1.0$, 
$(\pi\pi)_1$ etc. pairs $\alpha_V^e(\{8\}_a)=1.0, \alpha_V^m(\{8\}_a)=0.387$, 
$(\pi\rho)_1$ etc. pairs $\alpha_A(\{8\}_a)=0.652$. 

Unlike in \cite{RS96a,RS96b}, we did not fix pair couplings using
a theoretical model, based on heavy-meson saturation and chiral-symmetry.
So, in addition to the 14 parameters used in \cite{RS96a,RS96b} we now have
6 pair-coupling fit parameters. 
In Table~\ref{tab.gpair} the fitted pair-couplings are given.
Note that the $(\pi\pi)_0$-pair coupling gets contributions from the $\{1\}$ and
the $\{8_s\}$ pairs as well, giving in total $g_{(\pi\pi)}=0.10$, which has the
same sign as in \cite{RS96b}. The $f_{(\pi\pi)_1}$-pair coupling has opposite
sign as compared to \cite{RS96b}. In a model with a more complex and realistic
meson-dynamics \cite{SR97} this coupling is predicted as found in the present 
ESC-fit. The $(\pi\rho)_1$-coupling agrees nicely with $A_1$-saturation, see 
\cite{RS96b}. We conclude that the pair-couplings are in general not well
understood, and deserve more study.

The ESC-model described here is fully consistent with {\rm SU(3)}-symmetry. 
For the full {\rm SU(3)} contents of the pair interaction Hamiltonians we refer to
paper II, section III. Here, one finds for example that 
$g_{(\pi\rho)_1} = g_{A_8VP}$, and
besides $(\pi\rho)$-pairs one sees also that $(K K^*(I=1)$- and 
$K K^*(I=0)$-pairs contribute to the {\it NN} potentials.
All $F/(F+D)$-ratio's are taken fixed with heavy-meson saturation in mind, 
which implies that these ratios are $0.4$ or $1.0$ depending on the heavy-meson
type.
The approximation we have made in this paper is to neglect the baryon mass
differences, i.e. we put $m_\Lambda = m_\Sigma = m_N$. This because we
have not yet worked out the formulas for the inclusion of these mass 
differences, which is straightforward in principle.

\section{ Discussion and Conclusions }                             
\label{sec:9} 
We mentioned that we do not include negative energy state contributions.
It is assumed that a strong pair suppression is operative at low energies
in view of the composite nature of the nucleons. This leaves us for the
pseudoscalar mesons with two essential equivalent interactions: the 
direct and the derivative one. In expanding the $NN\pi$- etc. vertex in   
$1/M_N$ these two interactions differ in the $1/M^2_N$-terms, see \cite{RS96a} 
equations (3.4) and (3.5). This gives the possibility to use instead of the 
interaction in (\ref{eq:3.1}) the linear combination
\begin{eqnarray}
 {\cal H}_{ps} &=& \frac{1}{2}\left[(1-a_{PV}) g_{NN\pi} \bar{\psi}i\gamma_5
 \mbox{\boldmath $\tau$}\psi\cdot\mbox{\boldmath $\pi$} + 
 \right.\nonumber\\ && \left. 
 a_{PV} (f_{NN\pi}/m_\pi) \gamma_\mu\gamma_5\mbox{\boldmath $\tau$}\psi\cdot
 \partial^\mu\mbox{\boldmath $\pi$}\right]\ ,
\label{eq:6.1}\end{eqnarray}
where $g_{NN\pi}= (2M_N/m_\pi) f_{NN\pi}$. In ESC04 we have fixed $a_{PV}=1$, 
i.e. a purely derivative coupling.


The presented ESC-model is successful in describing the {\it NN}-data, 
even in this {\rm QPC}-constrained version. Allowing total freedom in the couplings 
and cut-off masses, and without fitting the low-energy parameters, 
we reached the lowest $\chi^2_{p.d.p.} = 1.10$. However, in that case some 
couplings look rather artificial. With some less freedom, a typical fit with
ESC-model has $\chi^2_{p.d.p.} = 1.15$, see e.g. \cite{RPN02}.
This means that by constraining the parameters rather strongly,
In the present {\it NN}-model ESC04 we we reached $\chi^2_{p.d.p.}= 1.155$, i.e. 
 we have only an extra $\Delta\chi^2 \approx 250$, showing the feasibility
of the {\rm QPC}-inspired couplings.

The gain of this is that we have physical motivated {\rm OBE}-couplings etc.. 
We will see in the 
next paper of this series, where we study the $S=-1$ {\it YN}-channels, 
that this feature persists when we fit {\it NN} and {\it YN} simultaneously. 
Then, the advantage is that going to the
$S=-2$ {\it YN}- and {\it YY}-channels, it is reasonable to believe that 
the predictions made for these channels are realistic ones. 
So far, there did not exist a 
realistic {\it NN}-model with sizeable axial-vector mesons couplings 
as predicted by Schwinger \cite{Sch68}.
Also, the zero in the scalar form factor has moderated the $f_0(760)$-coupling
such that it fits with the {\rm QPC}-model.

A momentum space version of ESC04 is readily available, using the material in 
\cite{RPN02}. We only have to add the momentum space potentials for the axial-vector
mesons, and the Graz-corrections \cite{Graz78}, which is rather straightforward.

Finally, the potentials of this paper are available on the Internet \cite{online}.



\appendix

\section{Axial-vector-meson coupling to nucleons}
\label{app:A}
The coupling of the axial mesons ($ J^{PC}= 1^{++}$) to the
nucleons is given by
\begin{eqnarray}
   {\cal L}_{ANN} &=&
 g_{A} \left[\bar{\psi} \gamma_{5}\gamma_{\mu}\bm{\tau}\psi\right]
     \cdot{\bf A}^{\mu}  +  i \frac{f_{A}}{\cal M}
    \left[\bar{\psi} \gamma_{5}\bm{\tau}\psi\right]
     \cdot\partial_{\mu}{\bf A}^{\mu} \nonumber\\
 &\approx & g_{A} \left[\bar{\psi} \gamma_{5}\gamma_{\mu}\bm{\tau}\psi\right]
     \cdot{\bf A}^{\mu}
\label{eq:A.1} \end{eqnarray}
Here, ${\cal M}= 1$ GeV is again a scaling mass.       
We note that with $f_{A}=0$ this coupling is part of the
$A_{1}$-interaction to pions and nucleons
\begin{equation}
   {\cal L}_{I} =  2 g_{A} \left[\bar{\psi}
\gamma_{5}\gamma_{\mu}\frac{1}{2}\bm{\tau}\psi +
 \left(\bm{\pi}\partial_{\mu}\sigma-\sigma\partial_{\mu}\bm{\pi}
 \right) + f_{\pi}\partial_{\mu}\bm{\pi}
\right] \cdot{\bf A}^{\mu}\ , \nonumber \\
\label{eq:A.2}\end{equation}
which is such that the $A_{1}$ couples to an almost conserved
axial current (PCAC). Therefore, the $A_{1}$-coupling used here is
compatible with broken ${\rm SU(2)}_{V}\times {\rm SU(2)}_{A}$-symmetry, see e.g. 
\cite{Schw69,Alf73}.
For a more complete discussion of the $A_1$-couplings to baryons we refer 
to \cite{SR97}. The latter reveals that as far as the axial-nucleon-nucleon 
coupling is concerned it is indeed of the type indicated above.
 
\noindent In the Proca-formalism, for the axial-vector propagator enters 
the polarization-sum
\begin{equation}
   \Pi^{\mu\nu}(k)= \sum_\lambda \epsilon^{\mu}(k,\lambda) 
  \epsilon^{\nu}(k,\lambda)  = 
   -\eta^{\mu\nu}+ k^{\mu}k^{\nu}/m^{2}
\label{eq:A.3}\end{equation}
where $m$ denotes the mass of the axial meson and $\epsilon^\mu(k)$ the 
polarization vector.
Because
\begin{equation}
   \left[\bar{\psi}\gamma_{5}\gamma_{\mu}\psi\right] k^{\mu}k^{\nu}
   \left[\bar{\psi}\gamma_{5}\gamma_{\nu}\psi\right] =
   \left[-i\bar{\psi}\gamma_{5}\gamma_{\mu}k^{\mu}\psi\right]
   \left[+i\bar{\psi}\gamma_{5}\gamma_{\nu}k^{\nu}\psi\right]
\label{eq:A.4}\end{equation}
the second term in the 'propagator' gives potentials which exactly
are of the form as those of pseudo-vector exchange. 
We note that these $\Gamma_5(p',p)=\gamma_5\gamma\cdot k$-factors come from 
the $\partial^\mu$-derivative of the pseudo-vector baryon-current.
Then,        
\begin{eqnarray}
 && \bar{u}(p') \Gamma_5(p',p) u(p) \approx
 i\left[\mbox{\boldmath $\sigma$} \cdot({\bf p-p'}) 
 \right.\nonumber\\ && \left.
 \mp \frac{E({\bf p})-E({\bf p}')}{2M}\
 \mbox{\boldmath $\sigma$}\cdot ({\bf p+p'}) \right]
\label{eq:A.5}\end{eqnarray}
in contrast to what is used in \cite{Rij91}, where
in the $1/M$-term $\omega({\bf k})$ is taken, instead of the baryon
energy difference. 
Notice that the second term in (\ref{eq:A.5}) is of
order $1/M^2$ and moreover vanishes on energy-shell. Hence this term
we neglect. 
We write  
\begin{equation}
 \tilde{V}_{A} = \tilde{V}_{A}^{(1)} +
 \tilde{V}_{A}^{(2)}\ ,
\label{eq:A.6}\end{equation}
 where $ \tilde{V}_{A}^{(2)} = \tilde{V}_{PV}$
 with $ f_{PV}^{2}/ m_{\pi}^{2}  \rightarrow g_{A}^{2}/ m^{2} $.       
 The transformation to the Lippmann-Schwinger equation implies
 the potential
 \begin{equation}
    \tilde{\cal V}_{A} \cong  \left(1 - \frac{{\bf k}^{2}}{8M'M}
    -\frac{{\bf q}^{2}}{2M'M}\right) \tilde{V}_{A}
 \label{eq:A.7}\end{equation}
Below, $M' = M_N$ and $M=M_Y$, which are the average nucleon mass or an 
average hyperon mass, depending on the baryon-baryon system.
 
\subsection{ ${\cal V}^{(1)}_{A}$-potential term}     
\label{app:Aa}
Restriction to terms which are at most of order $1/M^{2}$, we
find for the potential in Pauli-spinor space for the
Lippmann-Schwinger equation for $\tilde{\cal V}_{A}^{(1)}$
Note here that, especially for the anti-spin-orbit term, that $(M,\bm{\sigma}_1)$ 
and $(M',\bm{\sigma}_2)$ go with line 1 respectively with line 2.
Defining
\begin{equation}
 {\bf k} = {\bf p'} - {\bf p}\ ,\
 {\bf q} = \frac{1}{2}({\bf p'} + {\bf p})\ ,
 \label{eq:A.8}\end{equation}
 and using moreover the approximation
\begin{equation}
   \frac{1}{M^{2}}+\frac{1}{M'^{2}} \approx \frac{2}{MM'}\ ,
 \label{eq:A.9}\end{equation}
the potential ${\cal V}_A^{(1)}$ is given in momentum space by   
\begin{eqnarray}
   \tilde{\cal V}_{A}^{(1)} &=& -g_{A}^{2}\left[
  \left(1+\frac{({\bf q}^{2}+{\bf k}^2/4)}{6M'M}\right)
   \bm{\sigma}_{1}\cdot\bm{\sigma}_{2}
 \right. \nonumber\\ && \left.
   + \frac{2}{M M'} \left(
  (\bm{\sigma}_{1}\cdot{\bf q})(\bm{\sigma}_{2}\cdot{\bf q})
  -\frac{1}{3}{\bf q}^2 \bm{\sigma}_{1}\cdot\bm{\sigma}_{2}\right)
 \right.\nonumber\\ && \left. 
   - \frac{1}{4M' M}\left(
  (\bm{\sigma}_{1}\cdot{\bf k})(\bm{\sigma}_{2}\cdot{\bf k})   
  -\frac{1}{3}{\bf k}^2 \bm{\sigma}_{1}\cdot\bm{\sigma}_{2}\right)
 \right. \nonumber\\ && \left.
   +\left(\frac{1}{4M^{2}}-\frac{1}{4M'^{2}}\right)\cdot   
    \frac{i}{2}(\bm{\sigma}_{1}-\bm{\sigma}_{2})
   \cdot{\bf q}\times{\bf k} 
   \right. \nonumber \\ & & \left. 
   + \frac{i}{4M' M} (\bm{\sigma}_{1}+\bm{\sigma}_{2})
  \cdot{\bf q}\times{\bf k} 
  \right]\cdot \left(\frac{1}{\omega^{2}}\right)\ ,
 \label{eq:A.10}\end{eqnarray}
Now, for a 
complete treatment one has to deal with the non-local tensor. 
Although this can be done, see notes on non-local tensor potentials 
\cite{Rij98}, in this work we use an approximate treatment. 
We neglect the purely non-local tensor potential by making 
in (\ref{eq:A.10}) the substitution
\begin{eqnarray}
  && \frac{1}{M M'} \left(
  (\bm{\sigma}_{1}\cdot{\bf q})(\bm{\sigma}_{2}\cdot{\bf q})
  -\frac{1}{3}{\bf q}^2 \bm{\sigma}_{1}\cdot\bm{\sigma}_{2}\right)
 \rightarrow \nonumber\\
  && -\frac{1}{4M M'} \left(
  (\bm{\sigma}_{1}\cdot{\bf k})(\bm{\sigma}_{2}\cdot{\bf k})
  -\frac{1}{3}{\bf k}^2 \bm{\sigma}_{1}\cdot\bm{\sigma}_{2}\right)
 \label{eq:A.11}\end{eqnarray}
leading to a potential with only a non-local spin-spin term.
With this approximation, (\ref{eq:A.10}) becomes
\begin{eqnarray}
   \tilde{\cal V}_{A}^{(1)} &=& -g_{A}^{2}\left[
  \left(1+\frac{({\bf q}^{2}+{\bf k}^2/4)}{6M'M}\right)
   \bm{\sigma}_{1}\cdot\bm{\sigma}_{2}
 \right. \nonumber\\ && \left.
   - \frac{3}{4M' M}\left(
  (\bm{\sigma}_{1}\cdot{\bf k})(\bm{\sigma}_{2}\cdot{\bf k})   
  -\frac{1}{3}{\bf k}^2 \bm{\sigma}_{1}\cdot\bm{\sigma}_{2}\right)
 \right. \nonumber\\ && \left.
   +\left(\frac{1}{4M^{2}}-\frac{1}{4M'^{2}}\right)\cdot   
    \frac{i}{2}(\bm{\sigma}_{1}-\bm{\sigma}_{2})
   \cdot{\bf q}\times{\bf k} 
   \right. \nonumber \\ & & \left. \hspace*{0mm}
   + \frac{i}{4M' M} (\bm{\sigma}_{1}+\bm{\sigma}_{2})
  \cdot{\bf q}\times{\bf k} 
 \right]\cdot \left(\frac{1}{\omega^{2}}\right)\ .
 \label{eq:A.12}\end{eqnarray}
Then, we find in configuration space
\begin{eqnarray}
 && {\cal V}_{A}^{(1)} = - \frac{g_{A}^{2}}{4\pi}\ m  \left[
 \phi^{0}_{C}(m,r) (\bm{\sigma}_{1}\cdot\bm{\sigma}_{2}) \right. \nonumber\\
   && \left. -\frac{1}{12M'M}
  \left( \nabla^{2} \phi^{0}_{C}+\phi^{0}_{C}\nabla^{2}\right)(m,r)
 (\bm{\sigma}_{1}\cdot\bm{\sigma}_{2})
 \right. \nonumber \\ && \nonumber \\  && \left. \hspace*{0.0cm}
   + \frac{3m^{2}}{4M' M}\ \phi^{0}_{T}(m,r)\ S_{12}
 +\frac{m^{2}}{2M'M}\ \phi^{0}_{SO}(m,r)\ {\bf L}\cdot{\bf S}
 \right. \nonumber \\ && \nonumber \\ && \left. \hspace*{0.0cm}
 + \frac{m^2}{4M'M}\frac{M^{\prime 2}-M^2}{M'M}\ \phi^{(0)}_{SO}(m,r)\cdot
    \frac{1}{2}(\mbox{\boldmath $\sigma$}_1-\mbox{\boldmath $\sigma$}_2)\cdot{\bf L}
  \right]\ . \nonumber\\
 \label{eq:A.13}\end{eqnarray}

\subsection{ ${\cal V}^{(2)}_{A}$-potential term}     
\label{app:Ab}
For the PV-type contributions we have \cite{Graz78} 
\begin{eqnarray}
   \tilde{\cal V}_{A}^{(2)} &=&-\frac{g_{A}^{2}}{m^{2}}
   \left(1 - \frac{{\bf k}^{2}}{8M'M}-\frac{{\bf q}^{2}}{2M'M}\right)
 \cdot \nonumber\\ && \times 
  (\mbox{\boldmath $\sigma$}_1\cdot{\bf k})(\mbox{\boldmath $\sigma$}_2\cdot{\bf k})\
 \left(\frac{1}{\omega^{2}}\right)\ . 
\label{eq:A.14}\end{eqnarray}
The corresponding potentials in configuration space are
\begin{eqnarray}
   {\cal V}_{A}^{(2)} &=&  \frac{g_{A}^{2}}{4\pi}\ m\  \left[
  \frac{1}{3}(\mbox{\boldmath $\sigma$}_1\cdot\mbox{\boldmath $\sigma$}_2)
  \phi^{1}_{C} +\frac{1}{12M'M} 
 (\mbox{\boldmath $\sigma$}_1\cdot\mbox{\boldmath $\sigma$}_2)
 \right. \nonumber\\ && \left.
  \left( \nabla^{2} \phi^{1}_{C}+ \phi^{1}_{C}\nabla^{2} \right) +
  S_{12}\ \phi^{0}_{T} 
 \right. \nonumber \\ & & \left. 
  +\frac{1}{4M'M} \left( \nabla^{2} \phi^{0}_{T} S_{12} +
  \phi^{0}_{T} S_{12} \nabla^{2} \right)
 \right]\ ,
\label{eq:A.15}\end{eqnarray}

\acknowledgments
Discussions with prof. R.A. Bryan, J.J. de Swart, R.G.E. Timmermans, 
and drs. G. Erkol and M. Rentmeester are gratefully acknowledged.


\end{document}